\documentclass{aa}  
\usepackage{graphicx}
\usepackage{txfonts}
\usepackage{hyperref}
\hypersetup{
    colorlinks=true,
    linkcolor=cyan,
    citecolor=cyan,
    urlcolor=cyan,
    }

\makeatletter
\renewcommand*\aa@pageof{, page \thepage{} of \pageref*{LastPage}}
\makeatother

\begin{document}

   \title{Accretion-induced flickering variability among  \hbox{symbiotic stars from space photometry with NASA TESS}}


   \author{J.\,Merc\inst{\ref{inst:Prague}}\thanks{\email{jaroslav.merc@mff.cuni.cz}}
   \and P.\,G.\,Beck\inst{\ref{inst:IAC},\ref{inst:ULL}} 
\and S.\,Mathur\inst{\ref{inst:IAC},\ref{inst:ULL}} 
\and R.\,A.\,García\inst{\ref{inst:CEA}}
          }

   \institute{Astronomical Institute, Faculty of Mathematics and Physics, Charles University, V Holešovičkách 2, 180 00 Prague, Czechia\label{inst:Prague}
   \and Instituto de Astrof\'{\i}sica de Canarias, E-38200 La Laguna, Tenerife, Spain \label{inst:IAC}
\and Departamento de Astrof\'{\i}sica, Universidad de La Laguna, E-38206 La Laguna, Tenerife, Spain \label{inst:ULL}
\and Universit\'e Paris-Saclay, Universit\'e Paris Cit\'e, CEA, CNRS, AIM, 91191, Gif-sur-Yvette, France \label{inst:CEA}}

   \date{Received 29 September 2023 / Accepted 18 December 2023}

 
  \abstract
   {Symbiotic binaries exhibit a wide range of photometric variability spanning different timescales. These changes can be attributed to factors such as orbital motion, intrinsic variability of individual components, or the interaction between the two stars. In the range from minutes to hours, variability induced by accretion processes, likely originating from the accretion disks, denoted as flickering, is detected. This variability could mimic solar-like oscillations exhibited by luminous red giants.}
   {We aim to investigate whether it is possible to utilize the precise observations of the NASA TESS mission to detect flickering in symbiotic stars despite such studies being usually performed at shorter wavelengths than TESS observes. Additionally, our goal is to develop a quantitative method for the detection of accretion-induced flickering that does not rely solely on subjective assessment of the light curves.}
   {We obtain the light curves of known symbiotic stars and a comprehensive control sample of assumed single red giants from the TESS full-frame images. To ensure consistency, all the data are processed using the same methodology, which involves filtering out background, systematic, and long-term trends. From the processed light curves and their power spectral densities, we measure the amplitudes of the variability and other relevant parameters.}
   {We introduce a method that enables the differentiation between flickering sources and stars that do not exhibit this type of variability. We detect flickering-like variability in 20 symbiotic stars utilizing TESS data, with 13 of them being previously unidentified as flickering sources. Moreover, the TESS observations facilitate the detection of related variations occurring over timescales of a few days, as well as changes in the flickering behavior across multiple sectors.}
   {The flickering has now been likely detected in a total of 35 known symbiotic stars. Although this represents only a small subset of all symbiotic binaries, when focusing solely on accreting-only symbiotic stars where the detection of flickering is presumably more straightforward, the fraction could reach as high as $\sim$\,80\%. This suggests that accretion disks may be rather prevalent in these binaries.}


   \maketitle
%

\section{Introduction}\label{sec:intro}

\begin{table*}
\caption{Literature sample of confirmed flickering sources from the New Online Database of Symbiotic Variables.
}             
\label{tab:known_flickering}      
\centering          
\begin{tabular}{lrrll}
\hline\hline
Name & TIC & TESS $T$ {[}mag{]} & TESS sectors\,$^a$ & Flickering ref. \\
\hline      
ASAS\,J190559-2109.4 & 908998 & 10.30 & - & [1] \\
RS\,Oph & 6348255 & 9.47 & (80) & [2,3,4,5] \\
SU\,Lyn & 11670502 & 5.00 & 20, 60, (73) & [6] \\
V1044\,Cen & 30541762 & 8.69 & 37, 64 & [1] \\
BX\,Mon$^b$ & 64333488 & 8.61 & 7 & [5] \\
BF\,Cyg$^b$ & 109842110 & 9.09 & 14, 40, 41, (74, 81) & [5] \\
EG\,And$^b$ & 115780639 & 5.30 & 17, 57 & [5] \\
ASAS\,J152058-4519.7 & 146068950 & 9.16 & 38, 65 & [1] \\
Gaia\,DR2\,5917238398632196736 & 172960991 & 11.43 & 12, 39, 66 & [1] \\
Gaia\,DR2\,6043925532812301184 & 175694878 & 11.84 & 12, 39 & [1] \\
RT\,Cru & 180045004 & 8.88 & 11, 37, 38, 64, 65 & [7,8] \\
V2116\,Oph & 193746936 & 13.79 & - & [9] \\
CM\,Aql$^b$ & 308686857 & 11.55 & 54 & [5] \\
omi\,Cet (Mira) & 332890609 & 2.26 & 4, 31 & [5,10,11,12] \\
CH\,Cyg & 350739215 & 3.73 & 14, 15, 40, 54, 55, 56, (74, 75, 80, 81, 82) & [3,5,13,14] \\
V407\,Cyg & 357457104 & 9.52 & 15, 16, 55, 56, (75, 76, 82, 83) & [15,16] \\
CN\,Cha$^c$ & 394916057 & 7.02 & 11, 12, 38, 39, 64, 66 & [17] \\
V694\,Mon & 403675153 & 8.65 & 7 & [3,5,16,18,19] \\
EF\,Aql & 439643508 & 10.74 & 54, (81) & [20] \\
ZZ\,CMi & 453173127 & 6.38 & 7, 33 & [21] \\
T\,CrB & 462607643 & 7.73 & 24, 25, 51, (78) & [5,16,22,23] \\
V648\,Car & 463038319 & 6.93 & 9, 10, 36, 37, 63, 64 & [24]\\
\hline                  
\end{tabular}
\tablefoot{$^a$\,TESS sectors up to 67 were available and analyzed in this work. Sectors for which data is not yet available or observations are planned are presented in brackets. $^b$\,The presence of flickering is not fully confirmed but is highly probable. $^c$\,The flickering detection is based only on the TESS data. References in the Table: 1: \citet{2021PhDT........17L}, 2: \citet{1977MNRAS.179..587W}, 3: \citet{1996AJ....111..414D}, 4: \citet{2018MNRAS.480.1363Z}, 5: \citet{2001MNRAS.326..553S}, 6: \citet{2023BlgAJ..38...83Z}, 7: \citet{1994A&AS..106..243C}, 8: \citet{2023A&A...670A..32P}, 9: \citet{1996IAUC.6489....1J}, 10: \citet{2010ApJ...723.1188S}, 11: \citet{2018MNRAS.477.4200S}, 12: \citet{2019BlgAJ..31..110Z}, 13: \citet{1968IBVS..291....1C}, 14: \citet{1978MNRAS.185..591S}, 15: \citet{2003ARep...47..777K}, 16: \citet{2006AcA....56...97G}, 17: \citet{2020AJ....160..125L}, 18: \citet{1984BAAS...16..516B}, 19: \citet{1993ApJ...409L..53M}, 20: \citet{2017AN....338..680Z}, 21: \citet{2021AN....342..952Z}, 22: \citet{1988AJ.....95.1505L}, 23: \citet{2016MNRAS.462.2695I}, 23: \citet{2012ApJ...756L..21A}.}
\end{table*}

Symbiotic stars are strongly interacting binaries with long orbital periods, containing an evolved red giant and a white dwarf (or occasionally a neutron star) embedded in a circumbinary nebula. They constitute unique astrophysical laboratories important for understanding the binary evolution and the various processes that also occur in many other types of astrophysical objects \citep[see the reviews by][]{2012BaltA..21....5M, 2019arXiv190901389M}. 

These binaries exhibit significant photometric and spectroscopic variability on timescales between minutes to years, stemming from orbital motion effects (such as reflection, ellipsoidal effects, and eclipses), intrinsic variability of individual components (such as pulsations of the giant, rotation, or oscillations of the hot component), and interactions between the two stars (including outbursts and accretion-induced variability. Symbiotic systems typically have orbital periods ranging from a few hundred days to tens of years \citep[the peak in the distribution is between 500 and 600 days; e.g.,][]{1999A&AS..137..473M,2012BaltA..21....5M,2013AcA....63..405G}. It was pointed out by \cite{Beck2023} that this orbital period range is significantly lower than the distribution maximum of non-interacting red giant binaries, which have their orbital distribution peak between 1\,000 and 2\,000\,days.  While most of these stars are red-clump stars, red-giant branch stars are found on orbits with periods compatible with symbiotic stars \citep{2022A&A...667A..31B,Beck2023}. The pulsations of the cool components occur on timescales of 50-200 days for semiregular pulsators \citep[e.g.,][]{2013AcA....63..405G} or 300-600 days for symbiotic Miras \citep[e.g.,][]{1987PASP...99..573W,2009AcA....59..169G}. The most prominent changes in the light curves arise from the interaction between the components, leading to outbursts lasting weeks (symbiotic recurrent novae), months to years (active stages of classical symbiotic stars), or even decades ('slow' symbiotic novae), see, e.g., \citet{2007BaltA..16....1M,2010arXiv1011.5657M,2019arXiv190901389M}.

In addition to these long-term variations, accretion-induced, stochastic photometric fluctuations known as '\textit{flickering}' (in the symbiotic community\footnote{Not to confuse with the granulation-driven light curve '\textit{flicker}' known in the asteroseismic community \citep[e.g.,][]{2013Natur.500..427B,2016ApJ...818...43B,2014A&A...570A..41K}.}) have been observed in the light curves of a few symbiotic stars, characterized by amplitudes of several hundredths or tenths of a magnitude and time scales ranging from minutes to hours \citep[see, e.g.,][]{1996AJ....111..414D, 2001MNRAS.326..553S,2006AcA....56...97G}. This variability exhibits an increasing amplitude toward the blue end of the spectrum and is most pronounced in the near-UV, while it is relatively negligible in the red part of the optical region, mainly due to the dominant contribution from the cool giant \citep[e.g.,][]{2013A&A...559A...6L,2016MNRAS.461L...1M,2016acps.confE..21S,2021MNRAS.505.6121M}.

The phenomenon of flickering is not limited to symbiotic stars but is prevalent in various other accreting systems, including young stellar objects, accreting white dwarfs in cataclysmic variables, neutron stars and stellar-mass black holes in X-ray binaries, as well as supermassive black holes in active galactic nuclei \citep[see, e.g., ][and references therein]{2002ApJ...572..392B,2015SciA....1E0686S}. The underlying mechanism causing the flickering signal in photometry is believed to be the accretion disk, which produces similar red-noise aperiodic variability, characterized by a broken power-law power spectral density (PSD) shape and a strong correlation between flickering amplitude and average flux \citep[so-called rms-flux relationship; see ][for the analysis of this relation in some symbiotic stars]{2015MNRAS.450.3958Z,2016MNRAS.457L..10Z}. The exact physical processes that occur in the disks are not yet fully understood, and their detailed discussion is beyond the scope of this work. Therefore, we only mention one of the promising models as an example, which is the fluctuating accretion disk model, in which variations in viscosity at different radii on the local viscous time scale lead to modulations in the accretion flow, causing the observed variability \citep[see, e.g., ][and references therein]{1997MNRAS.292..679L,2006MNRAS.367..801A,2013MNRAS.434.1476I,2014MNRAS.438.1233S}. 

In cataclysmic variables, accreting white dwarf systems in which the donor is a red dwarf filling its Roche lobe \citep[e.g.,][]{1995cvs..book.....W,2001cvs..book.....H}, the flickering is easily detectable\,\citep[e.g.,][]{1992A&A...266..237B,2021MNRAS.503..953B}, as the contribution of the accretion disk emission dominates the optical region (the luminosity of the red dwarf is rather small). Moreover, the accretion disks of cataclysmic systems can also be studied thanks to the emission lines in the optical spectra (in particular of hydrogen) that arise from the disk and have a typical double-peaked structure. In contrast, symbiotic stars exhibit strong emission lines arising from the circumbinary nebula that is not present in cataclysmic variables. These nebular emission lines would obscure any emission lines arising from the disks in symbiotic systems, rendering flickering, although difficult to detect, the main observable evidence and a direct probe of accretion disks in them. Consequently, analyzing flickering becomes crucial not only for understanding accretion in symbiotic stars and their disks but also for comprehending their activity, as certain models of classical symbiotic outbursts require the presence of accretion disks \citep[see, e.g.,][]{2002ASPC..261..645M,2003ASPC..303....9M,2006ApJ...636.1002S}.

To search for flickering in symbiotic stars from the ground, observations in the $B$ band are commonly used due to the limited sensitivity of modern CCD cameras at bluer wavelengths, where flickering amplitudes are higher \citep[e.g., in $U$; see][]{2021MNRAS.505.6121M}. Flickering detection typically relies on visual inspection of light curves or comparing the target's variability amplitude with nearby stars of similar brightness and colors. No consistent method to prove the existence of flickering detection has yet been established. So far, only 22 confirmed symbiotic stars that show (or at least highly likely show) flickering have been firmly identified, all within the Milky Way, with no extragalactic detections reported (see their list with references in Table\,\ref{tab:known_flickering}). It should be noted that this constitutes only a small fraction compared to the total number of known galactic symbiotic stars, currently at 283, listed in the New Online Database of Symbiotic Variables\footnote{\url{https://sirrah.troja.mff.cuni.cz/~merc/nodsv/}} \citep[NODSV;][]{2019RNAAS...3...28M,2019AN....340..598M}. The challenge lies in the small amplitude of optical flickering and the limitations of relatively small ground-based telescopes used for such searches. Furthermore, even in cases where flickering is undoubtedly detected, it may not be present at all epochs \citep[e.g., ][]{2020MNRAS.492.3107L,2021MNRAS.505.6121M,2022BlgAJ..37...62G,2023BlgAJ..38...83Z,2023A&A...670A..32P}. Therefore, it is necessary to obtain observations over an extended period of time.

In this study, we analyze the short-term variability of a large sample of known symbiotic stars using data from the NASA \textit{Transiting Exoplanet Survey Satellite} \citep[TESS;][]{2015JATIS...1a4003R} for the first time. Previous studies utilizing this precise space-based photometry, which overcomes some limitations of ground-based datasets while introducing others (discussed below), have only focused on a few individual symbiotic stars, namely CN\,Cha \citep{2020AJ....160..125L}, T\,CrB \citep{2023MNRAS.tmp..729S}, RT\,Cru \citep{2023A&A...670A..32P}, and IGR\,J16194-2810 \citep{2023A&A...676L...2L}. It's worth noting that subsequent to the completion of this study, preliminary analysis of TESS data for NQ\,Gem was reported by \citet{2023BAAA...64...59L} with a full analysis, including X-ray data, in preparation. At the same time, the authors mention possible hints of flickering in light curves of four other symbiotic stars, BD\,Cam, V1261\,Ori, V1044\,Cen, and V420\,Hya, however, they do not present the data and their analysis in detail. 

This paper is organized as follows. Section \ref{app:amplitudes} discusses the expected amplitudes of the flickering variability in the TESS passband. Section \ref{sec:obs_proc} presents the observational data used in this study, their limitations, processing, and analysis methods. Section \ref{sec:known_flickering} discusses the analysis results of TESS light curves for symbiotic systems with previously reported flickering, including a method to distinguish flickering variability from other contributions in the TESS data. Section \ref{sec:unknown_flickering} presents the results of applying the method to a sample of confirmed symbiotic stars without reported flickering and introduces newly detected flickering sources. Section \ref{sec:z_and} briefly discusses the periodic signals present in TESS data of some symbiotic stars. Finally, Section~\ref{sec:conclusions} presents our conclusions.

\begin{figure}
   \centering
   \includegraphics[width=1\columnwidth]{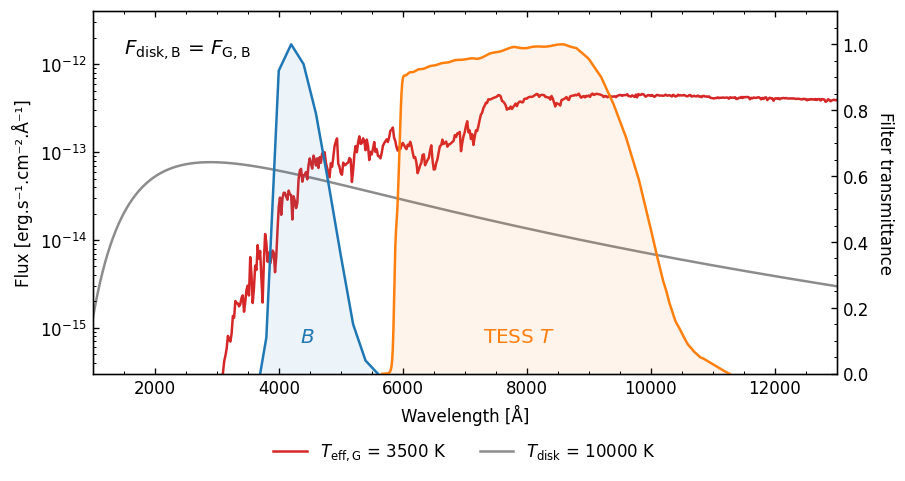}
      \caption{Visual representation of one of the scenarios discussed in Section \ref{app:amplitudes} where the flux of the red giant (depicted by the red line) and the accretion disk (represented by the black-body model in gray) are equal in the Johnson $B$ filter. The response curves of the Johnson $B$ and TESS $T$ filters are overplotted in blue and orange, respectively.}
         \label{fig:tess_amplitudes_B_vs_T}
\end{figure}

\section{Expected photometric amplitude of flickering variability in the TESS passband}\label{app:amplitudes}
The TESS mission captures observations in a specific TESS $T$ band, covering a wavelength range of approximately 600 - 1\,000 nm \citep[see the response function in Fig. \ref{fig:tess_amplitudes_B_vs_T};][]{2015JATIS...1a4003R}. This part of the spectral energy distribution of a symbiotic system is dominated by the luminous cool giant companion \citep[e.g.,][]{2005A&A...440..995S}, potentially obscuring accretion-induced variability associated with the disk.

\begin{figure}
   \centering
   \includegraphics[width=0.95\columnwidth]{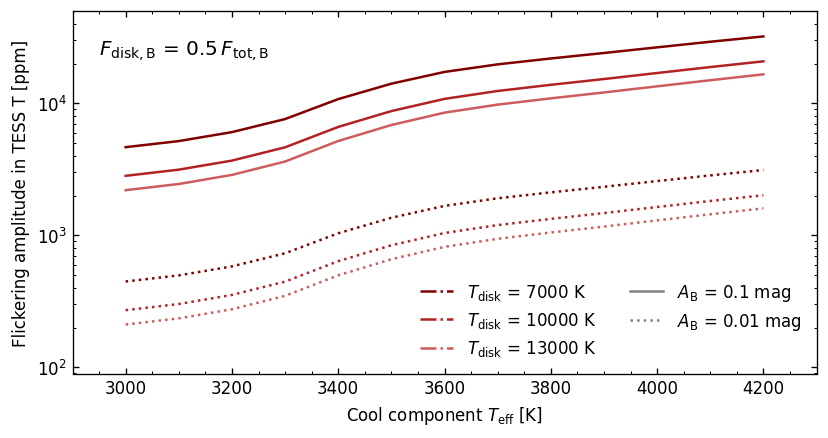}
      \caption{Expected flickering amplitude in the TESS $T$ band as a function of the effective temperature of the giant companion computed for three distinct accretion disk temperatures (7\,000 K, 10\,000 K, and 13\,000 K), assuming an equal integrated flux of the giant and the disk in the Johnson $B$ filter. The cases with flickering amplitudes, $A_{\rm B}$,  in the Johnson $B$ of 0.1 and 0.01 mag are illustrated by solid and dotted lines, respectively.}
         \label{fig:tess_amplitudes_disk_temperatures}
\end{figure}

\begin{figure}
   \centering
   \includegraphics[width=0.95\columnwidth]{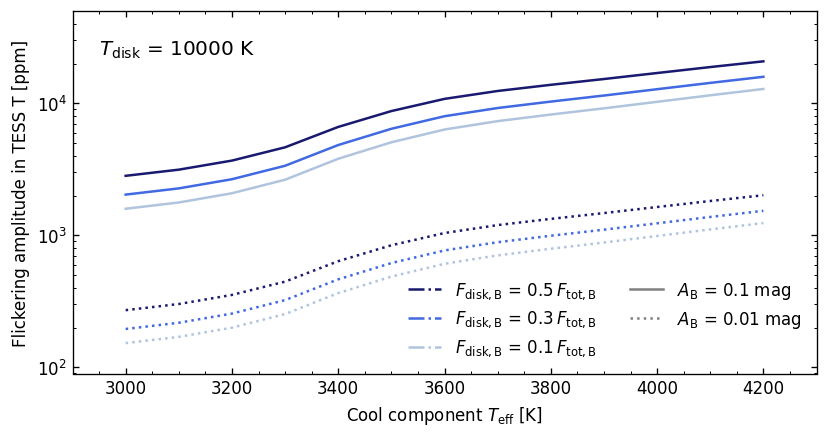}
      \caption{Same as Fig. \ref{fig:tess_amplitudes_disk_temperatures}, but calculated with a fixed disk temperature of 10\,000 K for three distinct ratios of disk to giant flux in the Johnson $B$ filter.}
         \label{fig:tess_amplitudes_flux_ratio}
\end{figure}

To assess the theoretical feasibility of detecting flickering in a symbiotic system with TESS, we estimated its amplitude in the TESS $T$ band, assuming that the observed amplitude in the Johnson $B$ filter falls within the range of 10-100 mmag. This range is typical for symbiotic stars, as reported in previous studies, based on ground-based photometry \citep[see, e.g., ][]{2001MNRAS.326..553S}. It is worth noting that, in some cases, even higher amplitudes have been observed. For instance, RS Oph exhibited a range of amplitudes from 0.16 to 0.59 mag in Johnson $B$ \citep{2018MNRAS.480.1363Z}. Similarly, CH Cyg displayed variability of up to 0.4 mag in observations by \citet{2018BlgAJ..28...42S}. V694 Mon, as reported by \citet{2020AN....341..430Z}, exhibited amplitudes ranging from 0.13 to 0.39 mag, while \citet{2023A&A...670A..32P} detected variability ranging from 0.23 to 0.37 mag in the case of RT Cru.

In the first scenario, we considered the optimistic case where the integrated flux of the accretion disk and the giant is equal in the Johnson $B$ filter. The radiation from the disk was modeled as a black body with three different temperatures (7\,000 K, 10\,000 K, and 13\,000 K), consistent with reported effective temperatures of flickering sources in symbiotic stars. For instance, \citet{2015MNRAS.450.3958Z} obtained a temperature of 9\,000 K for the flickering source in T CrB. \citet{2018BlgAJ..28...42S} reported a temperature range of 5\,000 - 11\,000 K for CH Cyg, \citet{2018MNRAS.480.1363Z} reported 7\,200 - 18\,900 K for RS Oph, and \citet{2020AN....341..430Z} obtained a temperature range of 6\,300 - 11\,000 K for the flickering source in V694 Mon. The radiation from the giant was modeled using the spectra from the BT-Settl grid \citep{2014IAUS..299..271A}. We computed the amplitudes for a range of giant effective temperatures spanning from 3\,000 to 4\,200 K in increments of 50 K. Finally, the Johnson $B$ and TESS $T$ filter profiles (Fig. \ref{fig:tess_amplitudes_B_vs_T}) employed in this analysis were acquired from the Spanish Virtual Observatory Filter Profile Service\footnote{\url{http://svo2.cab.inta-csic.es/theory/fps/}}\citep{2012ivoa.rept.1015R,2020sea..confE.182R}.

The outcomes of this analysis, as depicted in Fig. \ref{fig:tess_amplitudes_disk_temperatures}, revealed some general trends. The amplitude of the flickering variability in the TESS band tends to increase with higher giant temperatures, for the same Johnson $B$ filter amplitude. This is attributed to the lower ratio between the flux in the Johnson $B$ and the TESS $T$ band for the giant at the higher temperature end. Hence, if the flux of the giant and the disk is equal in the Johnson $B$ band, the discrepancy in flux between the giant and the disk in the TESS $T$ band would be less pronounced for hotter giants. Moreover, the amplitude of the flickering variability slightly rises with decreasing disk temperature, as the peak of its radiation shifts toward redder wavelengths. In all cases, the findings affirm that the anticipated flickering variability in TESS $T$ falls within the detectable range (approximately $10^2$-$10^4$ ppm).

In the second scenario, we maintained the disk temperature at 10\,000 K and examined the impact of varied flux ratios between the disk and the giant in the Johnson $B$ filter on the outcomes. We considered three cases where the disk contributed 10\%, 30\%, and 50\% of the total observed flux in Johnson $B$. The results, illustrated in Fig. \ref{fig:tess_amplitudes_flux_ratio}, indicate that although the amplitude tends to decrease with lower disk contributions, it generally remains within the detectable range in most instances.

To assess the validity of the predicted amplitudes, we can compare the Johnson $B$ and TESS $T$ flickering amplitudes observed nearly simultaneously in the case of RT Cru by \citet{2023A&A...670A..32P}. The comparison between our model and their observations reveals a commendable agreement, within the same order of magnitude.

\section{Observations, their limitations and processing}\label{sec:obs_proc}

\begin{figure}
   \centering
   \includegraphics[width=0.95\columnwidth]{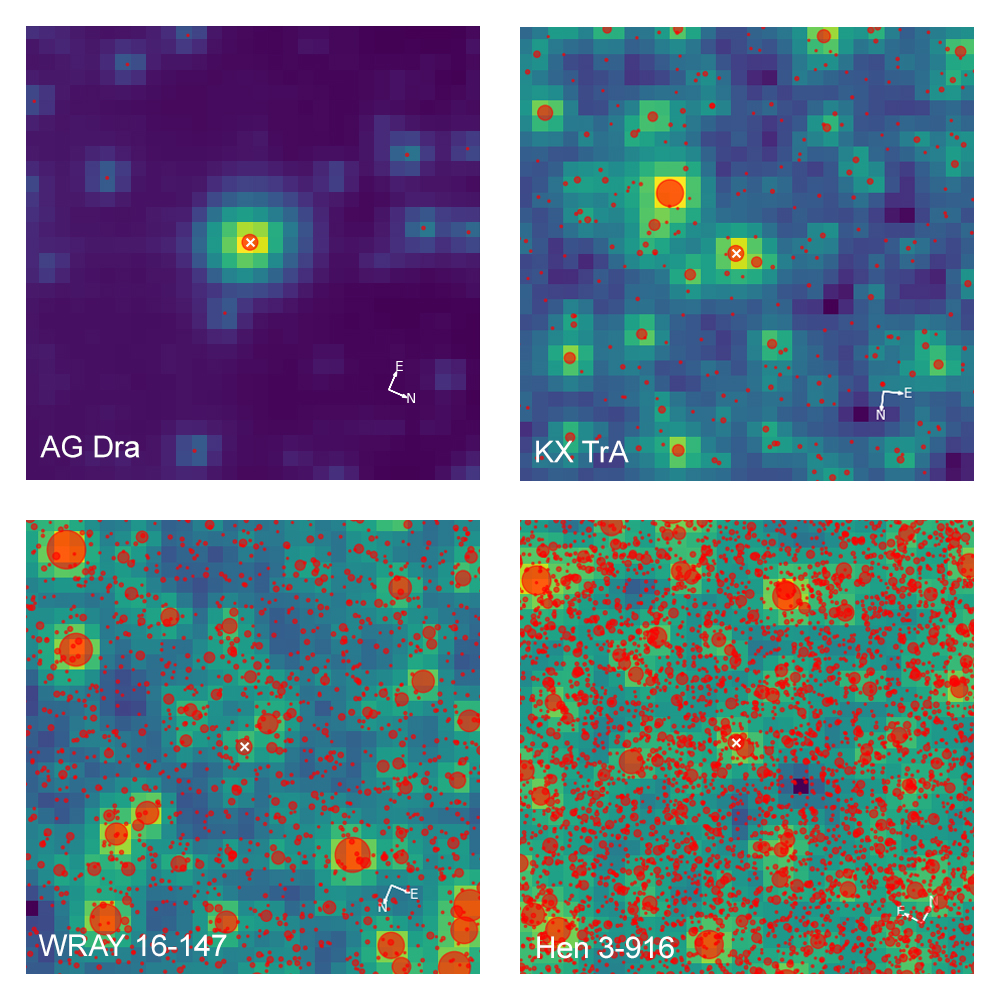}
      \caption{The target-pixel data of four symbiotic stars observed by TESS plotted using the \texttt{tpfplotter} tool \citep{2020A&A...635A.128A}. The top panel demonstrates examples of the least (AG\,Dra) and most contaminated fields (KX TrA) utilized in this study. The fields of WRAY\,16-147, with a galactic latitude of b~=~9.46$^{\circ}$, and Hen 3-916 (b~=-02.29$^{\circ}$), though not the most extreme cases, serve as examples of highly crowded regions that were excluded from our analysis. The fields shown cover 30\,x\,30 TESS pixels, corresponding to 10.5\,x\,10.5 arcminutes.}
         \label{fig:contamination}
\end{figure}

The TESS mission conducts an all-sky photometric survey by observing various portions of the sky in individual 'sectors', each covering a combined field of view of 24 by 96 degrees imaged by the 16 CCDs for approximately 27 days \citep[][]{2015JATIS...1a4003R}.  Due to the overlap between the sectors, some sky parts are observed continuously for an even longer time (up to almost one year in the so-called 'TESS continuous viewing zones' near the ecliptic poles). Individual 2-second exposures are combined into 2-minute images (with a 20-second mode introduced later during the first TESS extended mission). However, postage stamps of these data are available only for pre-selected targets, while full-frame images (FFIs) are downloaded to the ground with cadences of 30 minutes (Cycles 1 and 2; Sectors 1 -- 26), 10 minutes (first extended mission; Cycles 3 and 4; Sectors 27 -- 55), and 200 seconds in Cycle 5 and beyond (starting with Sector 56). The FFIs provided by the mission are calibrated, but the sky background (e.g., due to stray light from the Earth and Moon) is not removed and needs to be tackled during the analysis. 

\begin{figure}
   \centering
   \includegraphics[width=0.95\columnwidth]{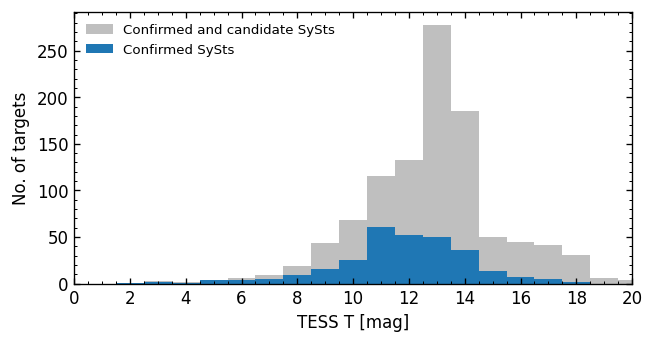}
      \caption{Distribution of TESS \textit{T} magnitudes of symbiotic stars and candidates from NODSV. In this work, we focused on confirmed symbiotic stars only (shown in blue).}
         \label{fig:magnitudes}
\end{figure}

\begin{figure}
   \centering
   \includegraphics[width=0.95\columnwidth]{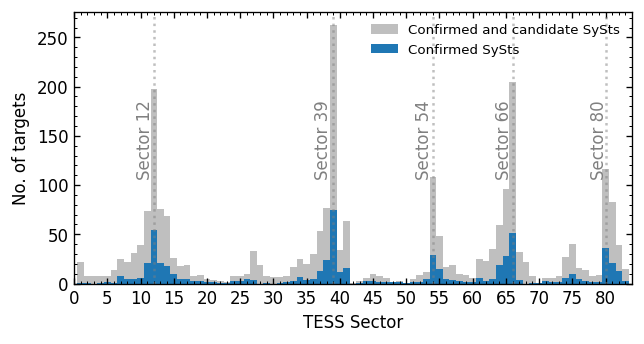}
      \caption{Distribution of confirmed symbiotic stars and symbiotic candidates from NODSV in the individual TESS sectors obtained using the \texttt{tess-point} tool \citep{2020ascl.soft03001B}. Sectors with the highest numbers of symbiotic stars are denoted by vertical dotted lines.}
         \label{fig:numbers}
\end{figure}

\begin{figure}
   \centering
   \includegraphics[width=0.95\columnwidth]{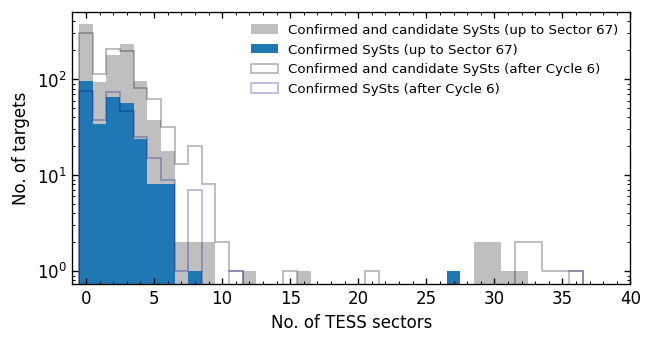}
      \caption{Number of TESS sectors available for the confirmed symbiotic stars and candidates. The shaded parts represent the sector availability at the time of writing, up to TESS Sector 67. The empty histogram indicates the projected availability after the completion of Cycle 6 observations in October 2024.}
         \label{fig:sectors}
\end{figure}

We should consider certain limitations of the TESS data that are relevant to our study. Probably most importantly, the pixel scale of the CCDs is 21 arcseconds per pixel, which poses challenges in crowded regions, such as the vicinity of the galactic plane where numerous symbiotic stars are located, due to contamination issues (see Fig.\,\ref{fig:contamination}). As previously noted in Section \ref{app:amplitudes}, the TESS wavelength range is predominantly influenced by the presence of the giant star, thereby reducing the amplitude of any potential accretion-induced variability. Additionally, the high brightness of symbiotic giants can lead to saturation in some cases, especially for stars brighter than a TESS \textit{T} magnitude of around 6. The excess charge is distributed along the CCD columns \citep{2021AJ....162..170H}, which can lead to issues for very bright targets when the bleed columns are too long. Although data for some of these objects may still have some utility, they are not ideal for detecting low-amplitude aperiodic changes on short timescales like flickering. Furthermore, the nature of the variability we are investigating imposes constraints on the target brightness, as faint targets may suffer from significant noise domination in their light curves. The reasonable upper limit for our purposes is about TESS \textit{T} $\sim$ 12 -- 13 mag. For fainter stars, the expected variability usually has a smaller amplitude than the noise level (see the discussion in Section \ref{app:amplitudes} and Sections \ref{sec:known_flickering} and \ref{sec:unknown_flickering}). The distribution of the TESS \textit{T} magnitudes of symbiotic targets (confirmed symbiotic stars and candidates) from NODSV is shown in Fig.\,\ref{fig:magnitudes}. 

The distribution of symbiotic stars and candidates in the individual TESS sectors is shown in Fig.\,\ref{fig:numbers}. Some symbiotic targets have been observed multiple times, resulting in data coverage spanning several sectors (Fig.\,\ref{fig:sectors}), although not necessarily continuously. It should be noted that these figures do not account for the contamination issue mentioned earlier and are not magnitude-limited, meaning that a considerable number of depicted symbiotic targets do not possess usable TESS light curves in reality. 

In this study, our primary focus was on the FFIs data, as the short cadence data were only available for a small subset of symbiotic stars (obtained within TESS Guest Investigator program G03206, PI: J. Merc, and as a by-product of other programs). However, if short cadence data, either 2-minute or 20-second cadence light curves processed by the TESS Science Processing Operations Center pipeline (SPOC) were available, we also examined them alongside our FFI light curves.

To extract the light curves from the FFIs, we employed the \texttt{Lightkurve} package \citep{2018ascl.soft12013L}. We downloaded the target pixel files, which consisted of a 30\,x\,30 pixel region centered on the target star, and performed aperture photometry to measure the flux of the object. A simple background subtraction technique was applied, where the median flux of the background pixel was subtracted from the flux measured within the aperture. To assess the extent of contamination affecting our target stars in the TESS data and identify sources with potentially unreliable light curves for flickering analysis, we conducted a thorough examination of their surroundings using the \texttt{tpfplotter} tool\footnote{\url{https://github.com/jlillo/tpfplotter}} developed by J. Lillo-Box \citep{2020A&A...635A.128A}. In this step, we considered all stars in the given field with magnitudes up to 6 mag fainter than the analyzed object. From the further analysis, we excluded symbiotic stars located near equally bright or brighter stars, as well as sources situated in crowded regions with multiple fainter neighboring sources. If there were only a few faint sources near the studied object, we examined its light curve and flagged them as 'possible contamination' in the tables and figures. Although it is highly unlikely that any periodic signal from nearby stars could be mistaken for flickering-like variability given our employed methods (such as the analysis of PSD, see below), in cases of potential contamination, we conducted a pixel-by-pixel analysis of TESS data to ensure, that the most probable source of the observed variability is the object of interest. In such a scenario, it is important to note that the observed amplitudes of the variability might be affected by the presence of neighboring objects within the photometric aperture.

\begin{figure}
   \centering
   \includegraphics[width=0.95\columnwidth]{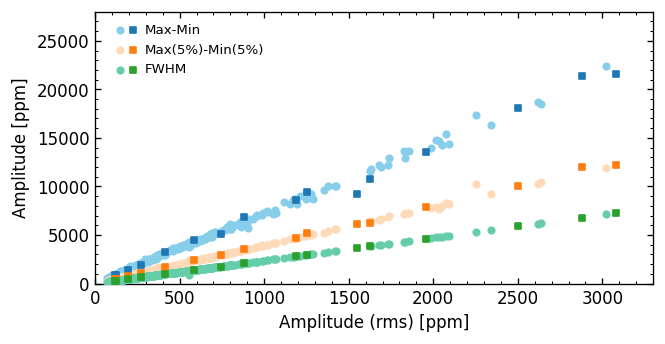}
      \caption{Variability amplitude obtained using different methods used in this study compared to the amplitude determined from the rms method (see the text for the details). The darker square and lighter circle symbols distinguish between symbiotic stars and red giants (see Sect.\,\ref{sec:known_flickering}), respectively.}
         \label{fig:amplitudes_comparison}
\end{figure}

To investigate variability on timescales ranging from minutes to hours, we applied various filtering techniques to the resulting light curves. Specifically, we utilized 1-, 3-, and 5-day triangular filters as well as the Savitzky-Golay (SG) filter \citep{1964AnaCh..36.1627S}. It is worth noting that the results obtained using different filters for smoothing yielded similar qualitative results. However, the application of wider triangular filters is not optimal due to the limited time base of the individual TESS sectors. Therefore, to avoid unnecessary repetition, only the results based on the SG filter are presented in this study. The filtering effectively eliminated long-term trends, whether real or systematic, from the light curves while preserving short-term variability (on the timescale of minutes and hours). At the same time, employing this approach helps to minimize the impact of potential inaccuracies in background subtractions on the results\footnote{Still some systematic effect might be preserved in the final data, such as the signal with a period of $\sim$3 days connected with the 'momentum dumps' of TESS. }. Additionally, we implemented sigma clipping (set to 4 $\sigma$ difference from a median in the particular filter window) to eliminate apparent outliers from the individual light curves. In instances where the artifacts in the light curves remained prominent even after the background removal process, we opted to utilize only the unaffected segments of the light curves. 

The resulting light curves were visually examined, and we measured the amplitudes of any possible variability using several methods commonly employed in the literature: a) peak-to-peak amplitude of variability \citep[difference between the maximum and the minimum in a light curve; e.g.,][]{2015MNRAS.450.3958Z,2016MNRAS.457L..10Z}; b) difference between the median of the top 5\% highest values and 5\% lowest value in a light curve; c) absolute root-mean-square (rms) amplitude of variability \citep{2005MNRAS.359..345U}; d) full width at half maximum (FWHM) of a Gaussian fit to the distribution of magnitude points in a light curve \citep{2021MNRAS.503..953B}. It is worth noting that if the light curves are filtered, and the expected value for the rms calculation is the same as the median, the latter two methods differ only by a constant. Additionally, our analysis of the light curves for symbiotic targets and single red giants demonstrated that all the employed methods yielded qualitatively very similar results (the absolute values are different, but the values are strongly correlated; Fig.\,\ref{fig:amplitudes_comparison}). Therefore, we present only the results obtained using the rms method. 
As described in the next section, we propose the Power Spectral Density (PSD) as an additional useful method to quantify accretion-induced flickering. We also compare these above-mentioned methods to the results of our new approach.



\section{Known flickering sources} \label{sec:known_flickering}
For the initial part of our analysis, we focus on confirmed symbiotic stars that have previously been detected showing flickering, as listed in Table\,\ref{tab:known_flickering}. However, not all of the 22 symbiotic stars in the list could be readily analyzed using TESS data. Two stars, namely ASAS\,J190559-2109.4 and V2116\,Oph, were neither observed in any of the available TESS sectors up to the present nor will they be in the planned Cycle 6 sectors. RS\,Oph, on the other hand, would only be observed in Sector 80 (June - July 2024), while data up to Sector 67 were available at the time of writing. Additionally, the observations of omi\,Cet (=\,Mira) and CH\,Cyg were found to be inapplicable for flickering analysis due to their excessive brightness ($T$ < 3 mag) causing detector saturation.

It is important to note that, additionally, several symbiotic stars from the list may have contaminated light curves. V1044\,Cen, for instance, is located very close to a bright eclipsing binary system \citep[ASASSN-V\,J131559.62-370018.8; P = 0.3828 d;][]{2018MNRAS.477.3145J}, which dominates the signal in the light curve. As a result, V1044\,Cen was excluded from further analysis. This leaves us with a total of 16 targets that were included in our study. However, it should be mentioned that Gaia\,DR2\,5917238398632196736, Gaia\,DR2\,6043925532812301184, CM\,Aql, and EF\,Aql are among the fainter stars in the list and are located in crowded regions. Additionally, there are some sources in close proximity to V407\,Cyg and RT\,Cru, although the star themselves are brighter than those mentioned previously. For this reason, special caution must be exercised when interpreting the results for these stars.

We acquired the TESS light curves for these targets, analyzing them on an individual sector basis to investigate temporal changes in variability (refer to Sect.\,\ref{sec:rt_cru}). Our analysis process involved an initial visual examination of the light curves, followed by calculations of the variability amplitude, PSD, and its fitting (see Sect.\,\ref{sec:obs_proc} for detailed information). To ensure our conclusions were not solely reliant on subjective assessments of the light curves, we aimed to establish a control sample of single red giants, as discussed in the following section.

\subsection{Distinction between flickering and stellar oscillations
}\label{sec:control}

\begin{figure}
   \centering
   \includegraphics[width=0.95\columnwidth]{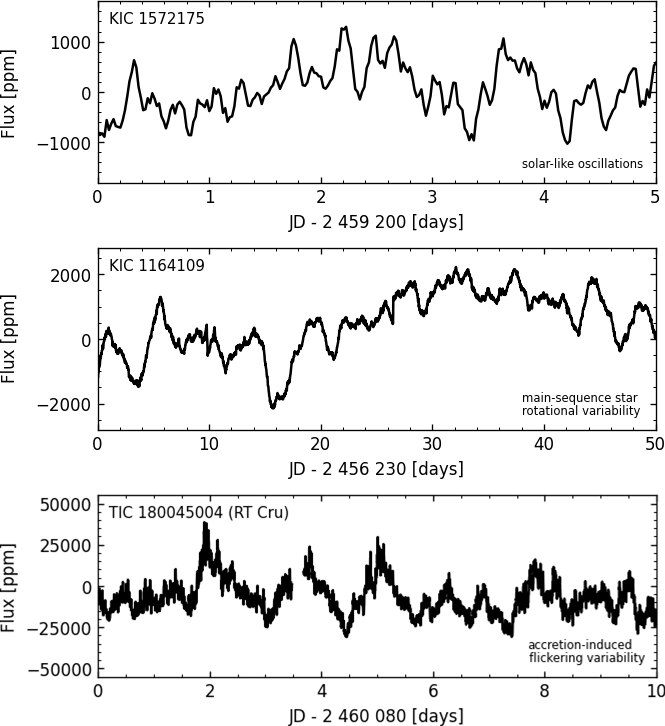}
    \caption{Comparison of the light curves of solar-like oscillating giant KIC\,1572175 (top panel), rotationally-modulated light curve of main-sequence star KIC\,1164109 (middle panel) and symbiotic star RT\,Cru exhibiting accretion-induced variability (bottom panel). The light curves of KIC\,1572175 and KIC\,1164109 were obtained by the \textit{Kepler} satellite and are taken from the KEPSEISMIC light curve database\protect\footnotemark\ \citep{2011MNRAS.414L...6G, 2014A&A...568A..10G,2015A&A...574A..18P}. The light curve of RT\,Cru is obtained from the TESS FFIs using the \texttt{Lightkurve} package (see Sect.\,\ref{sec:obs_proc}) and shows part of the Sector 65.}
         \label{fig:comparison}
\end{figure}

\footnotetext{\url{https://archive.stsci.edu/prepds/kepseismic/}}

Figure\,\ref{fig:comparison} illustrates the light curves of three distinct objects: TESS light curve of the symbiotic star RT\,Cru exhibiting accretion-induced flickering variability, and \textit{Kepler} light curves of the solar-like oscillating giant KIC\,1572175 \citep{2019MNRAS.485.5616H}, and the rotationally-modulated light curve of the main-sequence star KIC\,1164109 \citep{2022MNRAS.514.2793B}. Despite their different underlying mechanisms for variability, the light curves exhibit a remarkable resemblance. The apparent larger scatter in the RT\,Cru light curve is caused by the presence of the variability occurring on the shortest timescales of minutes.

Filtering out main-sequence stars from the potential sample of symbiotic stars, even when their variability timescales and light curve shapes are similar, is relatively straightforward, for instance, based on their positions in the \textit{Gaia} HR diagram \citep[see, e.g.,][]{2021MNRAS.506.4151M, 2022A&A...667A..31B}. However, distinguishing solar-like oscillations in luminous red giants \citep[see, e.g.,][]{2013A&A...559A.137M,2013ApJ...765L..41S,Mathur2016} from accretion-induced variability requires more careful consideration. The morphological similarity of the light curves poses a significant challenge in confirming flickering in symbiotic sources through visual evaluation of their light curves alone. As emphasized in Sect.\,\ref{sec:obs_proc}, the flux contribution of the cool component in the TESS band and its possible variability must be taken into account. In particular, cool giants might oscillate, and related processes, such as granulation, may contribute to the variability in their light curves significantly as well.

\begin{figure}
   \centering
   \includegraphics[width=0.95\columnwidth]{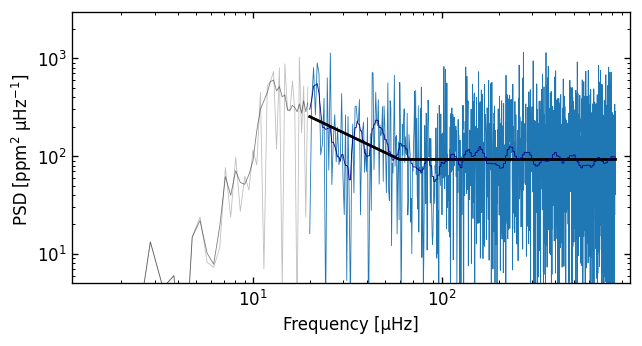}
      \caption{PSD of the Sector 27 data of the red giant TIC\,374859063, computed up to the Nyquist frequency. The input light curve was processed following the procedure described in Sect.\,\ref{sec:obs_proc}. The blue portion of the PSD was used in the subsequent analysis, where a piecewise linear fit in the log-log space was applied (black line). The gray part of the PSD is influenced by the SG filtering and is therefore excluded.}
         \label{fig:psd_example}
\end{figure}

The oscillating giant shown in Fig.\,\ref{fig:comparison} has a temperature $\sim$\,4\,900 K \citep[e.g.,][]{2022A&A...658A..91A}. This temperature is notably higher than the temperature range observed in the majority of symbiotic systems, where the peak of their distribution lies between 3\,200 and 3\,800 K \citep[e.g.,][]{2012BaltA..21....5M,2019RNAAS...3...28M,2019AN....340..598M,2019ApJS..240...21A}. We chose this example on purpose to demonstrate that the flickering could be masquerading as an asteroseismic signal. 
The timescales and amplitudes of oscillations and convection both rise as the star becomes more luminous \citep[][]{2011ApJ...743..143H,2012A&A...537A..30M,2014A&A...570A..41K} and the potential oscillations and granulation signal of cool giants, with temperatures typical of symbiotic cool components, are expected to manifest at longer timescales than those investigated in this study.
Figure\,1.3 in \citet{garcía_stello_2015} provides a depiction of typical oscillation PSD for stars across various temperature ranges.

The typical timescales of intrinsic variability in cool evolved stars motivated our selection of single giants as a reference sample for comparison with symbiotic sources. The comparison with randomly selected field red giants, despite being located nearby the target symbiotic stars in the sky, could introduce biases in the results. Therefore, we specifically chose \textit{Kepler} and TESS red giants from the catalogs of \citet{2020MNRAS.493.1388Y} and \citet{2021MNRAS.502.1947M}, with temperatures ranging from 3\,000 and 4\,000\,K, similar to the temperature range of the symbiotic star sample. The choice of the catalogs of oscillating sources not only ensured the inclusion of evolved red giants in our analysis, but the oscillation frequencies also provided independent confirmation of the correct temperature range. However, since the initial samples did not adequately cover the magnitude range below the TESS magnitude, \textit{T} = 8 mag, we supplemented them with additional red giants selected from the TESS Input Catalog \citep[TIC;][]{2019AJ....158..138S}. The selection from the TIC was based on various parameters, in particular, the magnitude, temperature, and luminosity class. In total, we included 335 supposedly single red giants in our control sample. The TESS light curves for all these stars were obtained and subjected to the same analysis as the symbiotic sources (see Sect.\,\ref{sec:obs_proc}). Upon visual examination of the resulting light curves, we confirmed that none of the analyzed red giants exhibited short-term flickering-like variability.

To allow the quantitative comparison, we computed the PSD of the light curves. A piecewise linear fit in the log-log space was applied to the smoothed PSD (‘logmedian’ method in \texttt{lightkurve}, that smooths the PSD using a moving median where the step size is determined by logarithmically increasing intervals in frequency space) of all the targets. From this fitting process, we obtained the low- and high-frequency slopes (in cases where higher frequencies were noise-dominated) or a single slope (when the noise level was not apparent from the PSD). Additionally, we determined the power at 25 and 215 $\mu$Hz as part of the analysis. Qualitatively, the PSD of all single red giants from our control sample look like the one shown in Fig.\,\ref{fig:psd_example}. The higher frequencies are dominated by the photon noise, leading to a frequency-independent background (as indicated by the 'high-frequency' slope close to 0 ppm$^2$\,$\mu$Hz$^{-2}$), whose amplitude is dependent on the brightness of the star. At the lower frequencies, some signal is detectable, as demonstrated by the non-zero slope of that part of the PSD. These are likely to be instrumental systematics. The typical frequency at which the slope changes is around 80 $\mu$Hz. 
We note that the 'low-frequency' slope is not very well constrained in our fitting procedure. This is caused by the fact that the length of the sectors analyzed is very limited ($\sim$27 days). As a result, the scatter in the PSD is rather large.

It is worth noting that the sample of red giants chosen for comparison with symbiotic sources exhibits some differences compared to giants in symbiotic systems. For example, single giants are not subjected to irradiation effects or tidal deformation. Nevertheless, these distinctions do not impact the results since the light curves of all giants in the control sample are noise-dominated at higher frequencies. Consequently, this sample effectively showcases the inherent noise properties in actual TESS observations, making it a more straightforward choice for comparison than relying solely on the noise performance presented in TESS documentation, especially considering the multitude of steps involved in our processing of TESS light curves.

Furthermore, \citet{2018A&A...612A..22B,2018MNRAS.479L.123B} have demonstrated that tidal interactions in binaries tend to suppress much of the intrinsic giant variability, except for activity-induced spot modulation of light curves. However, this phenomenon occurs with rotation periods that are usually synchronized with orbital periods in symbiotic stars \citep[see, e.g.,][]{2007MNRAS.380.1053Z} and are significantly longer than the timescales studied in this work.

\subsection{Detection of flickering with TESS}
As described in previous sections, we extracted a consistent set of parameters (specifically variability amplitudes, and slopes and power in PSD) to characterize the variability observed in both the TESS light curves of symbiotic stars and the control set of single red giants. For each target, we analyzed each sector individually, but when there were no significant changes in the behavior, we show the average values for all sectors in the resulting figures for clarity. By comparing these samples, we were able to explore effective quantitative methods for detecting flickering in symbiotic stars.

\begin{figure}
   \centering
   \includegraphics[width=0.95\columnwidth]{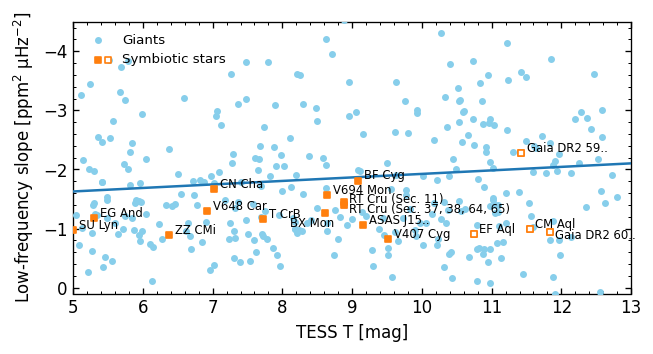}
      \caption{Low-frequency slope of the PSD of studied symbiotic sources (orange squares) and red giants from our control sample (blue circles) shown against the magnitudes of the targets from the TESS Input Catalog. Filled squares indicate isolated sources, while empty squares represent potentially contaminated sources. With the exception of RT\,Cru (Sect.\,\ref{sec:rt_cru}; two epochs are connected with a vertical line), the average values for all available sectors are shown in the figure. The blue line shows the formal linear fit to the dependence for giants.}
         \label{fig:low_frequency_slope}
\end{figure}

\begin{figure}
   \centering
   \includegraphics[width=0.95\columnwidth]{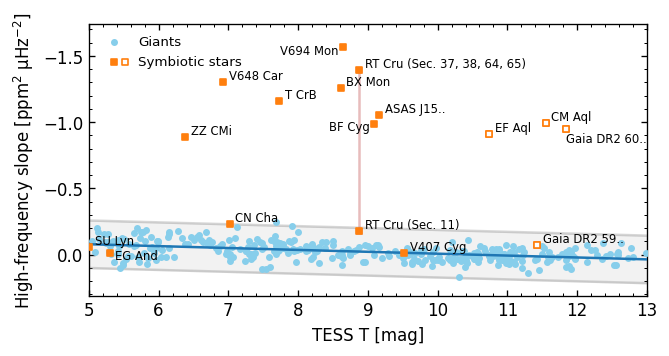}
      \caption{Same as Fig.\,\ref{fig:low_frequency_slope}, but for the high-frequency slope of the PSD. The shaded gray region shows the slopes $\pm$ 0.2 from the linear fit.}
         \label{fig:high_frequency_slope}
\end{figure}

\begin{figure}
   \centering
   \includegraphics[width=0.95\columnwidth]{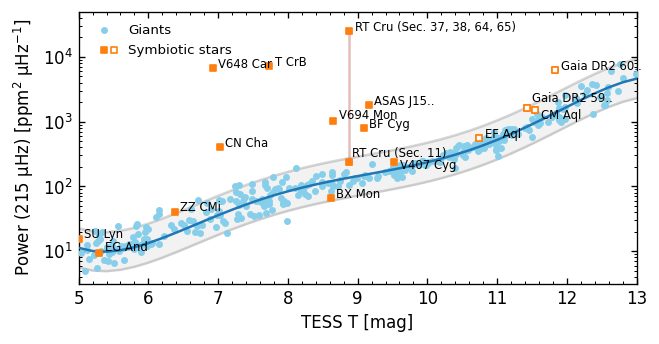}
      \caption{Same as Fig.\,\ref{fig:low_frequency_slope}, but for the power measured at the frequency of 215 $\mu$Hz. The blue line shows the polynomial fit to the dependence for the giants, and the gray lines are calculated as 0.5 and 2$\times$ the fit value for the given magnitude.}
         \label{fig:high_frequency_amplitude}
\end{figure}

\begin{figure}
   \centering
   \includegraphics[width=0.95\columnwidth]{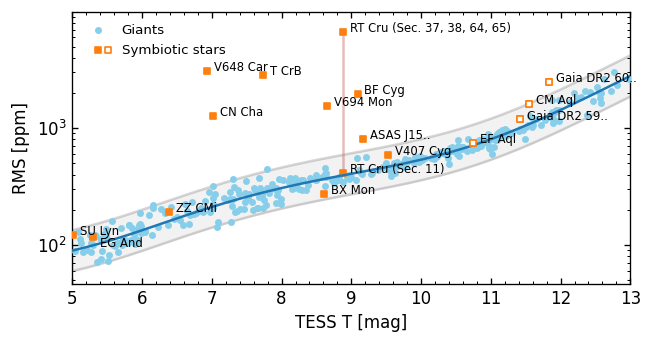}
      \caption{Same as Fig.\,\ref{fig:low_frequency_slope}, but for the rms variability amplitude. The blue line shows the polynomial fit to the dependence for the giants, and the gray lines are calculated as 2/3 and 1.5$\times$ the fit value for the given magnitude.}
         \label{fig:rms}
\end{figure}

In figures\,\ref{fig:low_frequency_slope}, \ref{fig:high_frequency_slope}, and \ref{fig:high_frequency_amplitude} we compare parameters of the PSD (slopes and power) for the group of confirmed flickering sources and our sample of single red giant stars. 
Figure\,\ref{fig:low_frequency_slope} reveals that the slope at low frequencies (typically $\nu$\,$\leq$\,80\,$\mu$Hz) is similar for both symbiotic stars and single red giants. Therefore, this portion of the PSD is not useful for detecting and characterizing flickering. Conversely, in Fig.\,\ref{fig:high_frequency_slope}, when comparing the slope at higher frequencies ($\nu$\,$\geq$\,80\,$\mu$Hz), all single red giants and some symbiotic sources exhibit a white noise-dominated behavior, reflected by a slope around zero. However, the figure also demonstrates that certain symbiotic stars exhibit some variability at these frequencies, as evidenced by their non-zero slope in this region of the PSD. 

An even more straightforward test of the flickering can be conducted by comparing the power in the high-frequency part of the PSD (Fig.\,\ref{fig:high_frequency_amplitude}). We specifically measured the power at 215 $\mu$Hz, either from the fit to the PSD or as a median in 10 $\mu$Hz wide box centered at that frequency (the latter is shown in Fig.\,\ref{fig:high_frequency_amplitude} as the results are virtually the same). At this frequency, all giants from the control sample within the studied magnitude range show only white noise (typically above $\sim$80 $\mu$Hz), but remains below the Nyquist frequency for the 30-minute cadence of the TESS observations ($\sim$278 $\mu$Hz). By selecting this frequency, we ensured that the parameter could be obtained consistently across all TESS sectors in which the targets were observed. Symbiotic stars exhibiting flickering-like variability in their light curves are clearly distinguishable in the diagram, appearing above the region where the single red giants are located. A similar conclusion can be drawn by analyzing the variability amplitudes measured as rms (Fig.\,\ref{fig:rms}). Unlike the power measured from the PSD, the rms method provides information about the overall variability in the light curve without distinguishing between different timescales. However, as demonstrated here, it can be effectively applied even to whole month-long light curves (corresponding to one sector), as long as any long-term variability is properly filtered out (SG filter in our case).

\begin{figure}
   \centering
   \includegraphics[width=0.95\columnwidth]{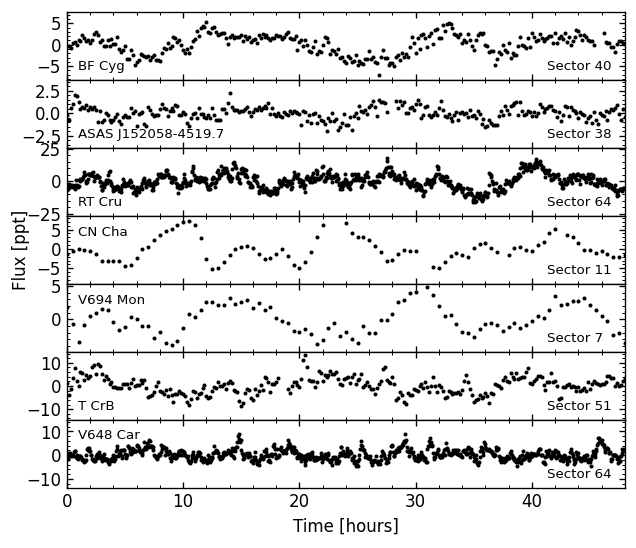}
      \caption{Two-days long parts of the SG filtered light curves of symbiotic stars exhibiting flickering in TESS with previous ground-based flickering detections.}
         \label{fig:hours}
\end{figure}

\begin{figure}
   \centering
   \includegraphics[width=0.95\columnwidth]{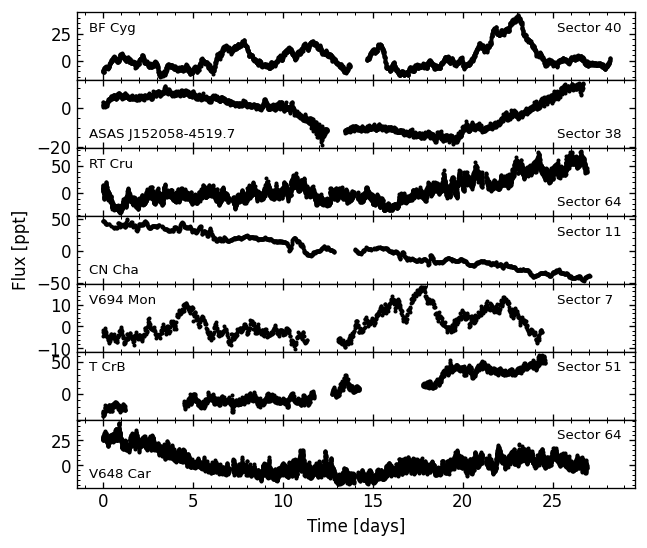}
      \caption{Unfiltered sector-long light curves of known flickering symbiotic stars exhibiting flickering in TESS. The figure demonstrates that these sources display variability not only on short timescales but also on longer timescales, likely associated with accretion processes (refer to the text for further details). The observed scatter in the light curves is a result of flickering occurring at minutes and hours timescales (see Fig.\,\ref{fig:hours}).}
         \label{fig:days}
\end{figure}

Based on the combined analysis shown in the diagrams in Figs. \ref{fig:high_frequency_slope}, \ref{fig:high_frequency_amplitude}, and \ref{fig:rms}, and the light curves (Fig.\,\ref{fig:hours}) we have determined that out of the 16 symbiotic stars studied, flickering-like variability can be reliably detected in seven of them: V648\,Car, CN\,Cha, T\,CrB, V694\,Mon, RT\,Cru, BF\,Cyg, and ASAS\,J152058-4519.7. We have also verified that nearby stars do not show similar variability using the same analysis as for the target stars. Among these targets, RT\,Cru is particularly intriguing due to its significant variability changes between sectors, which are discussed in more detail in Sect.\,\ref{sec:rt_cru}.

Intriguingly, our analysis of TESS observations of V694\,Mon obtained in Sector 7 (Jan 07, 2019 - Feb 02, 2019) has revealed signatures indicative of low-amplitude flickering-like variability. The star is presently undergoing a phase of unprecedented brightness, and since 2018, there has been a significant reduction in the amplitude of optical flickering of at least an order of magnitude when compared to earlier observations. Observations by \citet{2018ATel12227....1G}, \citet{2019ATel13236....1Z}, and \citet{2020ATel14239....1Z} did not identify any discernible flickering variability, establishing a limit of approximately 0.05 mag in the $B$ filter. A subsequent study by \citet{2021ATel15066....1M} reported a limit to the flickering amplitude of 0.005 mag during observations in November 2021. Remarkably, the star exhibited a nearly one-magnitude increase in brightness from October 2018 to November 2021 in $V$, and this upward trend persists in the latest observations. In stark contrast, previous observations had recorded amplitudes ranging from 0.13 to 0.39 mag \citep{2020AN....341..430Z}. The identification of low-amplitude variability through TESS observations does not invalidate the findings from ground-based studies, in particular that the observations were not taken simultaneously and the variability changes with time. Moreover, our model, discussed in Section \ref{app:amplitudes}, suggests that the amplitude in the $B$ band during the TESS observations likely falls below the reported limits.

Similarly to the case of V694\,Mon, we have identified flickering-like variability in T\,Crb, which was observed by TESS in Sectors 24-25 (Apr 16, 2020 - Jun 8, 2020) and Sector 51 (Apr 22, 2022 - May 18, 2022). In 2014, T\,Crb entered a super-active state that reached its peak in mid-2016 and recently concluded \citep{2023RNAAS...7..145M}. The super-active state is characterized by a notably reduced flickering amplitude in comparison to the quiescent variability. Ground-based observations in the $B$ filter by \citet{2016ATel.8675....1Z} revealed a flickering amplitude of 0.08 mag in February 2016. Subsequent observations from January to August 2023 detected variability in the range of 0.11 to 0.26 mag \citep{2023ATel15916....1S,2023ATel16023....1M,2023ATel16213....1Z}. This observed variability aligns broadly with the anticipated $B$ amplitudes inferred from the TESS observations. Our analysis, in conjunction with these data, confirms the presence of flickering in the super-active state, albeit at significantly lower levels compared to quiescence.

On the other hand, no evidence of flickering-like variability is observed in SU\,Lyn, EG\,And, ZZ\,CMi, BX\,Mon, V407\,Cyg, EF\,Aql, CM\,Aql, and Gaia\,DR2\,5917238398632196736. Lastly, Gaia\,DR2\,6043925532812301184 stands out slightly in the diagrams; however, due to its faintness and its location in a crowded region, the detection of flickering with TESS in this target remains uncertain. It is important to emphasize that the non-detection of flickering in these sources in our study simply indicates that flickering was not detected during the TESS observation, given the available precision, within the specific wavelength range observed by TESS. This does not invalidate the general detection of flickering reported in the literature for these sources, in particular, given that the expected amplitude of flickering in the TESS band is significantly lower, up to several orders of magnitude, compared to observations in the optical blue or near-UV (see Section \ref{app:amplitudes}). 

In addition to the study of variability on the timescales of minutes and hours, the relatively long duration of TESS sectors (approximately 27 days) allows for the investigation of brightness changes over extended periods with high cadence. Previous studies of brightness variations from one observing night to another have been limited in ground-based observations. Hence, we conducted a review of the unfiltered light curves of the sources exhibiting flickering in TESS. Interestingly, all stars exhibited brightness changes over the timescale of a few days, with larger amplitudes compared to shorter timescales (refer to their light curves in Fig.\,\ref{fig:days}). Among these, BF\,Cyg and V694\,Mon showed the most dramatic changes. The majority of this variability is likely associated with accretion processes, while the long-term trends may have different origins. For example, in the case of CN\,Cha, the trend reflects the recovery of the system from a 'slow' symbiotic nova outburst \citep{2020AJ....160..125L}, whereas in T\,CrB, the long-term trend spanning the sector is linked to the orbitally-related ellipsoidal effect \citep[e.g., ][]{2016MNRAS.462.2695I}.

\subsection{Changes in the flickering in RT\,Cru}\label{sec:rt_cru}
In most of the sources where flickering was detected with TESS, the parameters describing the variability remained relatively consistent across multiple sectors of observation. However, RT\,Cru is an exception to this pattern. Currently, data from four sectors are available for this star: Sector 11 (Apr 23, 2019 - May 10, 2019), Sectors 37 and 38 (Apr 02, 2021 - May 26, 2021), and Sectors 64 and 65 (Apr 06, 2023 - June 02, 2023).

\begin{figure}
   \centering
   \includegraphics[width=0.94\columnwidth]{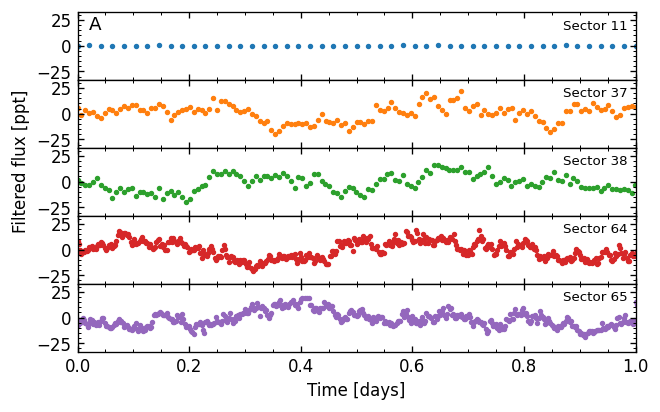}
   \includegraphics[width=0.91\columnwidth]{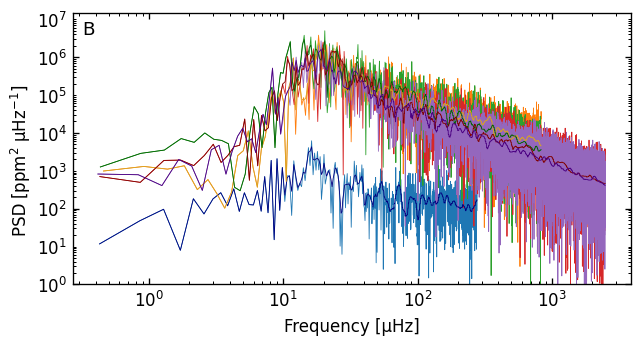}
      \caption{TESS data of the symbiotic star RT\,Cru. \textbf{A:} Processed and filtered light curves extracted from four observed sectors. Portions spanning one day are shown for clarity. \textbf{B:} PSD for the observed sectors, with each color representing a different light curve from panel A. 
      }
         \label{fig:rt_cru}
\end{figure}

During the observations in Sector 11, no significant variability was observed in RT\,Cru, and the star exhibited similar characteristics to single red giants in our control sample, as shown in Figs. \ref{fig:high_frequency_slope}, \ref{fig:high_frequency_amplitude}, and \ref{fig:rms}. However, in the three sectors observed subsequently (2 and 4 years later), strong flickering variability became evident (Fig.\,\ref{fig:rt_cru}). This change in behavior is not only noticeable when comparing with the control sample in the diagnostic diagrams but is also apparent in the light curves themselves (Fig.\,\ref{fig:rt_cru}A) and in the comparison of the PSD across individual sectors (Fig.\,\ref{fig:rt_cru}B). The substantial variation in flickering detected here with TESS was further confirmed by comprehensive ground-based follow-up observations, including photometric and spectroscopic measurements, as well as data from UV and X-ray observations \citep{2023A&A...670A..32P}. The authors of the study attributed the disappearance of flickering in 2019 to a decrease in the accretion rate, followed by a later recovery of the accretion flow through the disk in the subsequent years. 

Our analysis of RT\,Cru, depicted in Fig.\,\ref{fig:rt_cru}, serves as a compelling example, demonstrating the effectiveness of the methods developed in this work in distinguishing between sources with flickering and those without. It also highlights that the non-detection of flickering in some of the TESS observations does not definitively rule out the presence of this type of variability in a particular star, at least during certain epochs, and underscores the importance of repeated observations for analyzing the temporal variability in flickering.


\section{Search for new flickering sources in confirmed symbiotic systems}\label{sec:unknown_flickering}

\begin{figure*}
   \centering
   \includegraphics[width=0.92\textwidth]{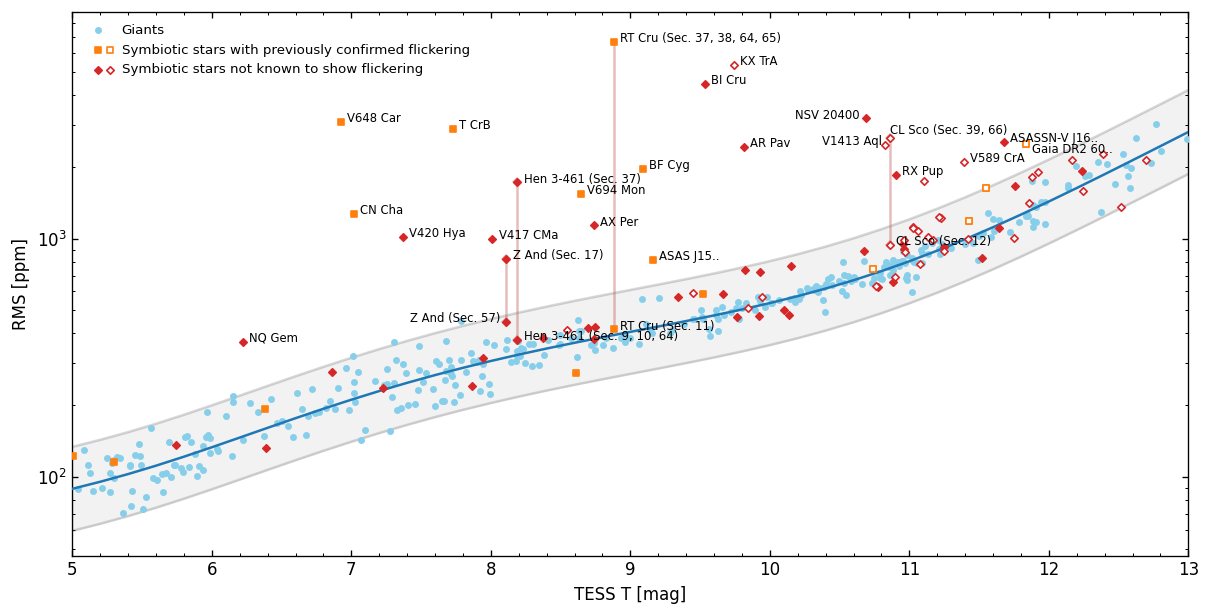}
      \caption{Diagram of the rms variability amplitude of symbiotic stars that were not previously known as flickering sources (depicted by red diamonds) alongside the known flickering sources (orange squares; Table \ref{tab:known_flickering} and Fig. \ref{fig:rms}) and red giants from our control sample (represented by blue circles). Filled and empty squares/diamonds indicate isolated and potentially contaminated sources, respectively. The average rms values are shown for sources that do not exhibit significant variation between sectors. }
         \label{fig:rms_unknown}
\end{figure*}

\begin{table*}
\caption{Newly classified flickering sources among symbiotic stars.}             
\label{tab:unknown_flickering}      
\centering          
\begin{tabular}{lrrlc}
\hline\hline
Name & TIC & TESS $T$ {[}mag{]} & TESS sectors$^a$ & Possible contamination \\
\hline      
Z\,And & 26011110 & 8.112 & 17, 57 &  \\
CL\,Sco & 34775634 & 10.861 & 12, 39, 66 & Yes \\
V417\,CMa & 63477133 & 8.012 & 6, 7, 33, 34 &  \\
NQ\,Gem$^b$ & 94748223 & 6.225 & 44, 45, 46, (71, 72) &  \\
RX\,Pup & 182401518 & 10.901 & 7, 8, 34, 35, 61, 62 &  \\
Hen\,3-461 & 303495210 & 8.191 & 9, 10, 36, 37, 63 &  \\
AR\,Pav & 304202168 & 9.816 & 13, 39, 66 &  \\
BI\,Cru & 311758230 & 9.534 & 11, 37, 38, 64, 65 &  \\
V1413\,Aql & 355716492 & 10.826 & 40, 53, 54, (80, 81) & Yes \\
NSV\,20400 & 442569144 & 10.689 & 12, 38, 65 &  \\
V420\,Hya & 443619246 & 7.367 & 10, 37, 64 &  \\
KX\,TrA & 447290900 & 9.74 & 12, 66 & Yes \\
AX\,Per & 453231036 & 8.741 & 18, 58 & \\
\hline                  
\end{tabular}
\tablefoot{$^a$TESS sectors up to 67 were available and analyzed in this work. $^b$Subsequent to the conclusion of this study, \citet{2023BAAA...64...59L} presented preliminary findings of flickering detection in NQ\,Gem using TESS data.}
\end{table*}

Our analysis, as described in the preceding section, has verified that despite the limitations of TESS data for studying flickering in symbiotic stars, there are cases where the amplitude is sufficient for detection in the relatively red band of TESS. Consequently, we have extended our analysis to include other confirmed symbiotic stars listed in NODSV in which the short-term variability either has not been yet studied or the flickering was not detected from the ground. Specifically, we focused on stars within the TESS \textit{T} magnitude range of 5 to 13 mag, which were observed in at least one TESS sector. Out of the 261 known symbiotic stars without flickering detection, 174 were located within the observed sectors up to Sector 67. Among these, 123 stars fell within the studied magnitude range. We excluded sources that were contaminated by nearby equally bright or brighter stars, as well as those situated in densely populated regions with numerous neighboring sources (within the photometric aperture). In total, we kept 72 targets for the analysis. Nonetheless, it is important to exercise caution when considering the possible detection of flickering in the remaining sources, as some may still be subject to contamination. When discussing the results in the subsequent analysis, we differentiate between sources that are well isolated in the TESS images and those that have other sources in close proximity, although these contaminants are fainter than the studied targets.

We inferred the same parameters of the 72 symbiotic stars as we did for the known flickering sources and the control sample of single red giants. The rms diagram for the studied targets is presented in Fig.\,\ref{fig:rms_unknown}. Most of the targets are located in the region of non-flickering single red giants (see their list in Table\,\ref{tab:unknown_nonflickering}). However, 15 objects exhibited a higher amplitude of variability than expected from white noise. The PSD power revealed a similar pattern, with some additional targets lying above the relation for the control sample. However, a careful study of the PSD of these stars revealed that the apparent higher power was caused by peaks or spikes in the PSD and not by the presence of flickering-like variability. 

\begin{figure}
   \centering
   \includegraphics[width=0.95\columnwidth]{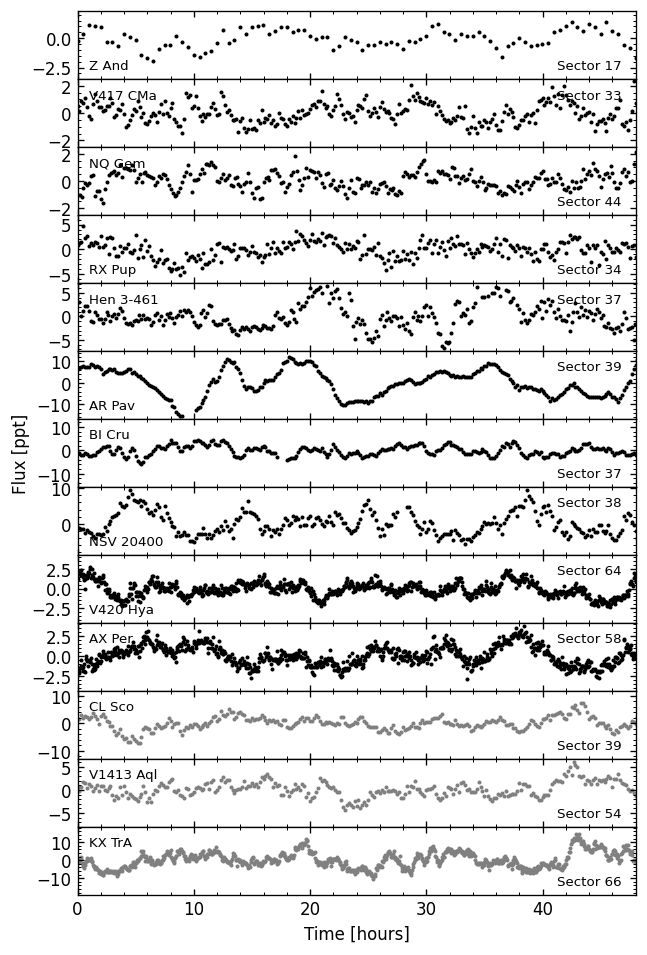}
      \caption{Two-days long parts of the SG filtered light curves of newly classified flickering sources among symbiotic stars. The black and gray light curves distinguish between isolated sources and potentially contaminated sources, respectively.}
         \label{fig:hours_unknown}
\end{figure}

\begin{figure}
   \centering
   \includegraphics[width=0.95\columnwidth]{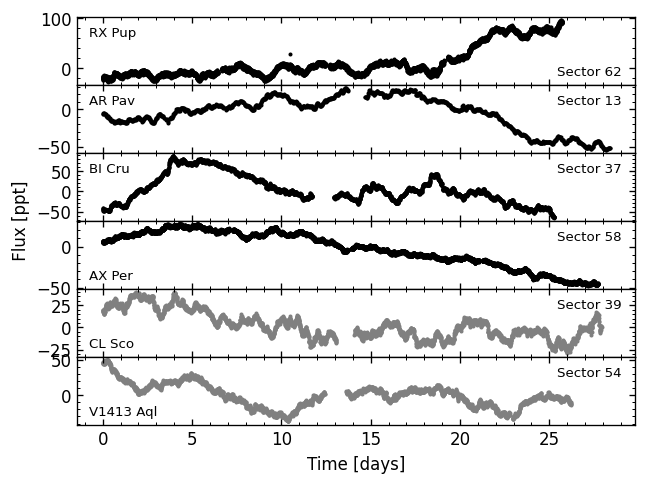}
      \caption{Unfiltered sector-long light curves of newly classified flickering sources that exhibit variability over several days. The colors of the light curves correspond to those in Fig.\,\ref{fig:hours_unknown}.}
         \label{fig:days_unknown}
\end{figure}

We thoroughly reviewed the individual light curves of all the stars, along with the light curves of comparison stars located nearby. In two cases, V589\,CrA and ASASSN-V\,J163807.84-284207.6, the light curves appeared to be affected by periodic signals (with periods of 2.57 h and 7.12 h, respectively; see Section \ref{sec:z_and}) and did not resemble flickering. Therefore, we do not classify these objects as flickering sources. The remaining 13 targets (listed in Table\,\ref{tab:unknown_flickering}) exhibited short-term variability reminiscent of flickering. Their light curves are shown in Fig.\,\ref{fig:hours_unknown}. Some of these sources also displayed variability on longer timescales ($\sim$\,days), as shown in Fig.\,\ref{fig:days_unknown}, similar to the previously known flickering sources discussed earlier.

Three of the symbiotic stars, namely Z\,And, Hen\,3-461, and CL\,Sco, exhibit flickering-like variability only in certain observed sectors, similar to the case of RT\,Cru (Sect.\,\ref{sec:rt_cru}). For Z\,And, the variability is detected in Sector 17 (Oct 08, 2019 - Nov 02, 2019), while three years later in Sector 57 (Sep 30, 2022 - Oct 29, 2022), this variability appears to be absent. Hen\,3-461 shows relatively constant behavior in Sectors 9 and 10 (Feb 28, 2019 - Apr 22, 2019), followed by prominent variability in Sector 37 (Apr 2, 2021 - Apr 28, 2021), and then returns to a low state in Sector 63 (Mar 10, 2023 - Apr 6, 2023). We note that the same pattern is visible also in 120 and 20-sec SPOC processed data, that are available for Hen\,3-461. Similarly, CL\,Sco exhibits a white-noise-dominated light curve in Sector 12 (May 21, 2019~-~Jun 18, 2019), while showing flickering in Sectors 39 (May 27, 2021 - Jun 24, 2021) and 66 (Jun 2, 2023 - Jul 1, 2023).

The appearance and disappearance of possible flickering in these sources may be connected to the activity of the systems. Z\,And experienced a very long active stage that began in 2000 \citep[e.g.,][]{2018ApJ...858..120S,2019OEJV..197...23M}, and the TESS observations were obtained during its decline phase, when the system may be transitioning to quiescence. The initial dataset for CL\,Sco was obtained during its low state before the outburst activity started in October 2019, as indicated by its ASAS-SN light curve \citep{2014ApJ...788...48S,2023arXiv230403791H}, and the system has since remained in a high state. Regarding Hen\,3-461, the detection of flickering occurred when the system was at its brightest over the past 7 years, as observed by the ASAS-SN survey. However, a more detailed analysis is necessary to determine if this brightness enhancement is associated with an outburst, as the light curves of Hen\,3-461 are complex due to the pulsations of the cool star and the orbitally-related variability.

Several of the newly detected flickering symbiotic stars had been previously studied from ground-based observations, but flickering was not detected at that time. Specifically, these stars include AR\,Pav \citep{1977MNRAS.179..587W}, AX\,Per \citep{2001MNRAS.326..553S}, and NQ\,Gem \citep{2012BlgAJ..18b..63S,2023BlgAJ..38...83Z}. In addition, RX\,Pup has been proposed as a possible recurrent nova with a Mira companion \citep{1999MNRAS.305..190M}. The authors suggested similarities between RX\,Pup and other symbiotic recurrent novae, such as RS\,Oph and T\,CrB. However, the detection of flickering was not available at that time to support their model. 


Finally, Z\,And has been sometimes previously included in lists of flickering sources, but its inclusion in those lists was based on the detection of a periodic short-term signal rather than aperiodic flickering variability \citep[see Sect. \ref{sec:z_and} and][]{1999ApJ...517..919S}.

\subsection{Occurrence-rate of flickering in symbiotic stars}

We detected flickering-like variability in TESS light curves of 20 confirmed symbiotic systems. For 13 sources, this type of variability is reported for the first time. This addition brings the total number of known symbiotic binaries with likely detection of flickering to 35, still accounting only for approximately 12\% of known galactic symbiotic stars in the NODSV.

However, we should highlight that this would be only a lower limit to the real number of symbiotic stars with flickering, as not all symbiotic binaries were observed by TESS, some sources are too faint to be subjected to thorough analysis using the available data, and densely populated regions introduce contamination issues that hinder drawing conclusive insights regarding any short-term variability. Furthermore, even when examining the subset of sources that had previously shown flickering, our analysis successfully detected some variability in only 7 out of 16 stars with available, relatively uncontaminated TESS data. This observation implies that the absence of flickering in TESS light curves does not necessarily indicate its absence altogether. It is plausible that repeated observations at different epochs or observations conducted at shorter wavelengths may unveil this variability in the stars where it was not apparent in the TESS data.

On top of that, previous studies on flickering in symbiotic binaries have suggested that in symbiotic systems, where hydrogen-rich material undergoes shell burning on the surface of the white dwarf \citep[referred to as shell-burning symbiotic stars; see, e.g., ][]{2019arXiv190901389M}, the luminosity generated by this process, in particular reprocessed to the optical by the symbiotic nebula, may be sufficiently high to overshadow the contribution from the accretion disk and potentially mask any flickering signal or at least significantly reduce its amplitude \citep[e.g.,][]{2001MNRAS.326..553S,2003ASPC..303..202S}. However, as demonstrated by shell-burning systems distinct from symbiotic stars, the process of the shell burning itself is not turning off the flickering variability, and these types of objects still can exhibit high-amplitude flickering if it is not diminished, e.g., due to the presence of a nebula. This phenomenon is evident in cases such as the recurrent nova T Pyx \citep{2001MNRAS.326..553S} or the super-soft X-ray binary MR Vel \citep{2003ASPC..303..202S}. It is important to emphasize that the flux from shell-burning alone remains relatively stable over short time scales, and the nebula itself is unlikely to introduce rapid variability  \citep{2003ASPC..303..202S}. Consequently, while the data analyzed in the current study may not be entirely conclusive in establishing a unique link between the observed variability and accretion processes, in particular for shell-burning systems, it appears likely.

More than two-thirds of the 35 likely flickering sources among symbiotic stars belong to the second, less prevalent group (because they are more challenging to detect) of accreting-only symbiotic stars that lack shell-burning. If we consider all accreting-only symbiotic stars from NODSV where flickering was previously detected through ground-based observations or those with usable TESS light curves (while excluding faint sources or those with contaminated light curves), the fraction of stars exhibiting detectable flickering rises to over 80\% (22 out of 27). Only five accreting-only symbiotic stars with usable TESS light curves did not display any flickering-like variability (V934 Her, UV Aur, ER Del, GSC 06806-00016, and GH Gem; Tab. \ref{tab:unknown_nonflickering}). These results, made possible by a substantial increase in the number of detected flickering sources in this study, strongly suggest that accretion disks are a common phenomenon in symbiotic stars.

\section{Period signal in Z\,And and in other symbiotic systems with TESS}\label{sec:z_and}
\begin{figure}
   \centering
   \includegraphics[width=0.95\columnwidth]{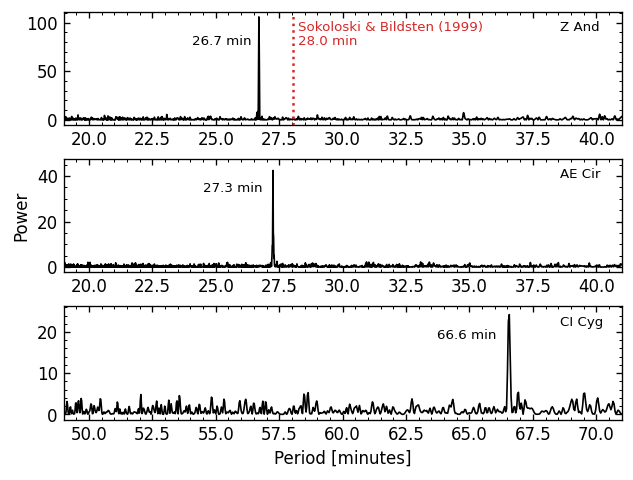}
      \caption{Lomb-Scargle periodograms of TESS light curves of Z\,And, AE\,Cir, and CI\,Cyg. The red dotted line in the upper panel represents the period inferred for Z\,And by \citet{1999ApJ...517..919S}.}
         \label{fig:periodograms}
\end{figure}

Z\,And has been known to show short-term periodic variability, distinct from the typical flickering, thanks to the observations of \citet{1999ApJ...517..919S}. The authors inferred a period of 1682.6 $\pm$ 0.6 s (approximately 28 minutes) from repeated \textit{B} band observations and attributed it to the rotation of an accreting magnetic white dwarf. The presence of a similar period was later confirmed by \citet{2006AcA....56...97G} through $U$ and $B$ band observations. In the current study, the TESS Sector 17 data lacked sufficient cadence (30 min) to study this periodicity. However, a single significant period was detected in the Sector 57 data (200 sec cadence) using a Lomb-Scargle method \citep[][]{1976Ap&SS..39..447L,1982ApJ...263..835S}, and its value was determined to be 1601.9 $\pm$ 0.6 s (26.7 min; Fig. \ref{fig:periodograms}). This suggests that the period has changed by approximately 80 seconds over a time span of 25 years. This change may be related to the ongoing prolonged activity of Z\,And, however, a detailed investigation of this periodic signal is beyond the scope of the current work and will be presented elsewhere.

It is worth mentioning that, in addition to Z\,And, we have also detected similar periodicities in the TESS light curves of two other objects: AE\,Cir with a period of 27.3 minutes and CI\,Cyg with a period of 66.6 minutes (Fig. \ref{fig:periodograms}). To increase the confidence that the identified variability is not attributed to neighboring sources, we utilized the \texttt{TESS-Localize} package \citep{2023AJ....165..141H}. This tool allows the localization of the variability source within the target pixel file, pinpointing the most probable star responsible for the observed variability. In the case of AE\,Cir, the association with the target star is highly probable, with the variability source located within 0.02 arcsec of AE\,Cir position. However, the case of CI\,Cyg is somewhat less definitive. While the signal is clearly discernible in the processed light curve, the localization within the target pixel file is less constrained. Notably, even though the \texttt{TESS-Localize} package identified the most probable source of the detected variability that coincides with CI\,Cyg, the positional constraints remain less definitive than desired. As expected, upon repeating the same analysis for Z\,And, the association of the variability source with the target star became undoubtful.

If the subsequent follow-up observations validate the inferred periodicities in AE\,Cir and CI\,Cyg as real and not related to any background sources, we could speculate that these periods may have a similar origin as the one observed in Z\,And. This discovery would significantly increase the sample of symbiotic stars thought to have magnetic white dwarfs, following Z\,And \citep{1999ApJ...517..919S} and FN\,Sgr, for which \citet{2023arXiv230605095M} inferred a rotation period of 11.3 minutes using \textit{Kepler} data. 

As mentioned earlier, periodic variability was also detected in the TESS light curves of V589\,CrA and ASASSN-V\,J163807.84-284207.6. However, the amplitudes and periods of these variations appear to be different from those observed in the TESS data of AE\,Cir, CI\,Cyg, and Z\,And. Furthermore, based on the \texttt{TESS-Localize} package analysis, the signal detected in the TESS light curve of V589\,CrA is undoubtedly linked to the pulsating variable star V756\,CrA, located approximately 1 arcmin away. In the case of ASASSN-V\,J163807.84-284207.6, the variable source seems to coincide with the studied star, and the origin of the 7.12-h variability remains unclear.

.

\section{Conclusions}\label{sec:conclusions}
In this study, we aimed to explore the short-term variability of symbiotic binaries using the precise photometric observations provided by the TESS space mission. Specifically, our focus was on the flickering phenomenon associated with the accretion disks around the hot components of these systems (white dwarfs or neutron stars). Flickering variability has been previously detected in only a small fraction of symbiotic systems, partly due to limitations imposed by ground-based observations, which are typically conducted in the optical region where the amplitude of flickering is significantly reduced compared to the near-UV range.

Although the rather red TESS passband is not ideally suited for studying this type of variability for the same reason, our findings support previous research indicating that flickering signatures can still be detected at the wavelengths observed by TESS, where the dominant radiation in symbiotic systems originates from the cool evolved giants. Since the flux contribution from the red giants is prominent, we conducted a detailed analysis of potential signals originating from these giant stars, including oscillations and granulation. We established a control sample consisting of presumed single red giants to facilitate a comparison with the symbiotic sources. By examining various parameters of the light curves, we aimed to identify the most effective indicators for distinguishing between sources that exhibit flickering and those that do not, thereby providing a quantitative method to complement the visual assessment of the light curves, that may sometimes be subjective.

Keeping in mind the limitations imposed by the large pixel scale of 21 arcseconds per pixel in crowded regions, we analyzed the TESS observations of both, already known flickering sources among the symbiotic stars and the confirmed symbiotic stars from NODSV for which the short-term variability either has not been studied or the flickering was not detected. Through our analysis, we were able to identify minutes-to-hours flickering-like variability in a total of 20 symbiotic stars, with 13 of them being previously unrecognized sources of flickering.

Currently, the number of known symbiotic stars showing likely accretion-induced variability on short timescales is 35. Though this constitutes only a small fraction of all known galactic symbiotic stars, our study demonstrated that if we consider only accreting-only symbiotic systems – in which flickering is presumably more easily detected – the fraction could be as high as around 80\%, possibly even higher. This finding strongly suggests that accretion disks are common in symbiotic systems.

In addition to detecting variability on timescales of minutes and hours, our analysis of nearly uninterrupted 27-day-long time series has revealed that the short-term flickering variability often correlates with changes occurring over a few days. By leveraging repeated observations across multiple sectors, we have observed fluctuations in the presence and amplitude of flickering over time. The phenomenon of flickering disappearing and reappearing has been previously documented in several sources, including the symbiotic star RT\,Cru, as observed by TESS. The inclusion of three additional symbiotic stars exhibiting this behavior in our sample of 20 TESS flickering sources suggests that such variability patterns are not uncommon in these systems. Furthermore, it is possible that comparable behavior may emerge in other systems as more TESS sectors become accessible in the future.

Finally, the findings presented in this study hold significant potential for future research, e.g., in near-UV using one of the planned UV facilities, such as Czech \emph{Quick Ultra-Violet Kilonova surveyor} which will focus, among others, also on symbiotic binaries as a secondary science objective \citep{2023arXiv230615080W,2023arXiv230615081K,2023arXiv230615082Z}. At the same time, our results serve as a precursor for the upcoming European space mission PLATO, scheduled for launch in 2026 \citep[]{2014ExA....38..249R,rauer_aerts_cabrera_plato}. 
The \textit{PLATO} mission is equipped with telescopes that provide a cadence of 25 seconds for stars fainter than 8 magnitudes. Additionally, it incorporates two multi-color fast cameras operating at a cadence of 2.5 seconds within the magnitude range of 4 to 8, and features a smaller pixel scale of 15 arcseconds per pixel. 
These improvements offer an excellent opportunity to further advance the study of accretion processes in symbiotic stars.\medskip


\begin{acknowledgements}
We are thankful to the referee, Ulisse Munari, for the comments and suggestions greatly improving the manuscript. JM acknowledges support from the Instituto de Astrofísica de Canarias (IAC) received through the IAC early-career visitor program and the support from the Erasmus+ programme of the European Union under grant number 2020-1-CZ01-KA203-078200. PGB acknowledges the support of the Spanish Ministry of Science and Innovation with the \textit{Ram{\'o}n\,y\,Cajal} fellowship number RYC-2021-033137-I and the number MRR4032204, and the financial support by \textit{NAWI\,Graz}. SM acknowledges the support of the Spanish Ministry of Science and Innovation with the \textit{Ram{\'o}n\,y\,Cajal} fellowship number RYC-2015-17697,  with the grants no. PID2019-107187GB-I00 and PID2019-107061GB-C66, and through AEI under the Severo Ochoa Centres of Excellence Programme 2020--2023 (CEX2019-000920-S). R.A.G. acknowledges the support from the PLATO and GOLF/SoHO Centre National D'{\'{E}}tudes Spatiales grant.

This research made use of the Spanish Virtual Observatory (https://svo.cab.inta-csic.es) project funded by MCIN/AEI/10.13039/501100011033/ through grant PID2020-112949GB-I00. This research made use of \texttt{lightkurve}, a Python package for Kepler and TESS data analysis \citep{2018ascl.soft12013L} and \texttt{tpfplotter} by J. Lillo-Box (publicly available in www.github.com/jlillo/tpfplotter).

Additional software used in this study includes \texttt{astropy} \citep{2013A&A...558A..33A,2018AJ....156..123A,2022ApJ...935..167A}, \texttt{matplotlib} \citep{Hunter:2007},
\texttt{NumPy} \citep{harris2020array}, and \texttt{Peranso} \citep{2016AN....337..239P}.

\end{acknowledgements}

%
\bibliographystyle{aa} 
\bibliography{bibliography.bib} 
%

\begin{appendix} 

\begin{table*}
\section{List of symbiotic stars with no flickering detection with TESS}
\caption{Studied symbiotic stars that have neither reported detection of the flickering in the literature nor the flickering was detected in their TESS light curves.}         \small    
\label{tab:unknown_nonflickering}      
\centering          
\begin{tabular}{lrrlc}
\hline\hline
Name & TIC & TESS $T$ {[}mag{]} & TESS sectors$^a$ & Possible contamination \\
\hline      
UV Aur & 2688613 & 6.387 & 19, 43, 44, 45, 59, (71, 73) &  \\
PN MaC 1-17 & 7064267 & 12.388 & 54, (80) & Yes \\
Hen 3-1213 & 26292371 & 9.825 & 12, 39, 66 &  \\
AE Ara & 30667659 & 9.929 & 12, 39, 66 &  \\
HK Sco & 34182969 & 11.134 & 12, 39, 66 & Yes \\
V455 Sco & 42538591 & 10.758 & 12, 66 & Yes \\
IV Vir & 46022938 & 9.34 & 11, 64 &  \\
V1922 Aql & 48042010 & 11.23 & 54 & Yes \\
RW Hya & 58761688 & 7.226 & 37, 64 &  \\
HD 319167 & 61299049 & 11.03 & 13 & Yes \\
CI Cyg & 83299618 & 7.941 & 14, 41, 54, 55, (74, 75, 81, 82) &  \\
StHA 149 & 84887747 & 9.764 & 26, 40, 53, (80) &  \\
V589 CrA & 90322314 & 11.39 & 13 & Yes \\
StHA 63 & 95247427 & 12.237 & 7, 34, 61 &  \\
GSC 06806-00016 & 98634059 & 10.139 & 12, 39 &  \\
Hen 2-375 & 118979030 & 11.64 & 13, 39, 66 &  \\
DD Mic & 126658353 & 9.925 & 1, 27, 67 &  \\
LAMOST J122804.90-014825.7 & 130588388 & 11.251 & 46 &  \\
Hen 3-160 & 139977336 & 12.516 & 8, 9, 34, 35, 36, 61, 62, 63 & Yes \\
WRAY 15-157 & 154230723 & 11.857 & 7, 8, 34, 61 & Yes \\
V1016 Cyg & 171876171 & 11.213 & 14, 15, 41, 54, 55, (75, 81, 82) & Yes \\
ASASSN-V J163807.84-284207.6 & 205521107 & 11.678 & 12, 39 &  \\
V347 Nor & 207551959 & 11.08 & 12, 39, 65, 66 & Yes \\
WRAY 16-145 & 210451236 & 11.418 & 11, 38, 65 & Yes \\
V1329 Cyg & 230861869 & 10.964 & 15, 41, 55, (75, 82) &  \\
AG Dra & 237100139 & 8.375 & 14, 15, 16, 17, 18, 19, 20, 21, 23, 24, &  \\ & & &
25, 26, 40, 41, 47, 48, 49, 50, 51, 52, &  \\ & & &
53, 54, 55, 56, 57, 58, 60, (74, 75, 76, &  \\ & & &
77, 78, 79, 80, 81, 82, 83) &  \\
V471 Per & 241112109 & 11.924 & 18, 58 & Yes \\
Hen 3-860 & 244876801 & 11.877 & 37, 38, 64 & Yes \\
Hen 3-863 & 244906894 & 10.957 & 11, 37, 64 &  \\
StHA 180 & 248105480 & 11.025 & 54, (81) &  \\
V934 Her & 257212213 & 5.745 & 25, 52, (79) &  \\
AS 255 & 261036112 & 11.172 & 13, 39, 66 & Yes \\
GH Gem & 261877049 & 12.163 & 33, 44, 45, (71, 72) & Yes \\
StHA 169 & 273371736 & 11.755 & 14, 15, 41, 55, (74, 75, 81, 82) &  \\
HD 330036 & 279627924 & 10.151 & 12, 39, 65 &  \\
ER Del & 282619434 & 7.866 & 55, (81) &  \\
StHA 32 & 298812903 & 11.524 & 5, 32 &  \\
SkySy 1-8 & 326902832 & 12.245 & 13, 39, 66 & Yes \\
YY Her & 329668061 & 10.779 & 26, 40, (80) &  \\
AS 276 & 331137325 & 11.10 & 66 & Yes \\
V443 Her & 333326188 & 8.75 & 26, 40, 53, (80) &  \\
V835 Cen & 333861396 & 12.698 & 11, 12, 38, 65 & Yes \\
V417 Cen & 334771334 & 9.448 & 11, 12, 38, 65 & Yes \\
HM Sge & 342021932 & 11.246 & 14, 40, 41, 54, (81) & Yes \\
V426 Sge & 345493023 & 9.847 & 14, 41, 54, (81) & Yes \\
Hen 3-1768 & 346727757 & 10.102 & 12, 13, 27, 39, 66, 67 &  \\
StHA 154 & 346758038 & 8.545 & 26, 40, 53, 54, (80) & Yes \\
QW Sge & 361557261 & 10.898 & 14, 41, 54, (81) & Yes \\
ASASSN-V J175047.54-390117.8 & 368084095 & 10.97 & 39, 66 & Yes \\
AG Peg & 385053976 & 6.859 & 55, (82) &  \\
StHA 190 & 388775434 & 9.667 & 55, (82) &  \\
TX CVn & 389852937 & 8.737 & 22, 49, (76) &  \\
WRAY 15-1470 & 392171429 & 11.06 & 12, 39 & Yes \\
PU Vul & 402762666 & 9.942 & 14, 54, 55, (81, 82) & Yes \\
LT Del & 406544192 & 11.751 & 14, 41, 55, (81, 82) & Yes \\
SS73 96 & 409145402 & 10.961 & 39, 66 & Yes \\
RR Tel & 421990485 & 10.884 & 13, 27, 67 &  \\
V399 Pav & 466125998 & 8.698 & 13, 27, 66, 67 &  \\
V366 Car & 469253499 & 10.675 & 9, 10, 36, 37, 63, 64 &  \\
\hline                  
\end{tabular}
\tablefoot{$^a$TESS sectors up to 67 were available and analyzed in this work. }
\end{table*}

\end{appendix}

\end{document}